\documentclass[preprint]{elsarticle}

\journal{Physica D}

\usepackage{amssymb}

\usepackage{amsfonts}
\usepackage{epsfig}
\usepackage{psfrag}
\usepackage{amsmath}
\usepackage{amssymb,subfigure}

\newcommand{\x}{{\bf x}}
\newcommand{\e}{{\bf e}}

\usepackage{color}

\definecolor{rred}{rgb}{0.7,0,0.1}
\definecolor{greenrb}{rgb}{0.2,0.6,0.3}
\definecolor{darkcyan}{rgb}{0.0,0.55,0.55}
\definecolor{ccyan}{rgb}{0.5,0,0.4}
\definecolor{orange}{rgb}{1.,0.5,0.}
\definecolor{purple}{rgb}{0.3,0.7,0}

\usepackage[normalem]{ulem} 

\begin{document}

\begin{frontmatter}

\title{Low-frequency variability and heat transport in a low-order nonlinear coupled ocean-atmosphere model}


\author[address1]{St\'ephane Vannitsem \corref{mycorrespondingauthor}}
\cortext[mycorrespondingauthor]{St\'ephane Vannitsem}
\ead{Stephane.Vannitsem@meteo.be}

\author[address1]{Jonathan Demaeyer}

\author[address1]{Lesley De Cruz}

\author[address2,address3]{Michael Ghil}

\address[address1]{Institut Royal M\'et\'eorologique de Belgique, Avenue Circulaire, 3, 1180 Brussels, Belgium}

\address[address2]{{Geosciences Department and Laboratoire de M\'et\'eorologie Dynamique (CNRS and IPSL), Ecole Normale Sup\'erieure, F-75231 Paris Cedex 05, France}}

\address[address3]{Department of Atmospheric \& Oceanic Sciences and Institute of Geophysics \& Planetary Physics, 405 Hilgard Ave., Box 951565, 7127 Math. Sciences Bldg., University of California, Los Angeles, CA 90095-1565, U.S.A.}

\begin{abstract}
We formulate and study a low-order nonlinear coupled ocean--atmosphere model with an emphasis on the impact of radiative and heat fluxes and
of the frictional coupling between the two components. This model version extends a previous  
24-variable version by adding a dynamical equation for the passive advection of temperature in 
the ocean, together with an energy balance model. 

The bifurcation analysis and the numerical integration of the model reveal the presence of low-frequency variability 
(LFV) concentrated on and near a 
long-periodic, attracting orbit. This orbit combines atmospheric and oceanic modes, and 
it arises for large values of the meridional gradient of radiative input and of frictional coupling. Chaotic behavior develops
around this orbit as it loses its stability; this behavior is still dominated by the LFV on decadal and multi-decadal time scales that is typical of oceanic processes. 
Atmospheric diagnostics also reveals the presence of predominant low- and high-pressure zones, as well as of a subtropical jet; 
 these features recall realistic climatological properties of the oceanic atmosphere.

Finally, a predictability analysis is performed. Once the decadal-scale periodic orbits develop, the coupled system's short-term instabilities  
--- as measured by its Lyapunov exponents --- are drastically reduced, indicating the ocean's stabilizing role  
on the atmospheric dynamics. On decadal time scales, the recurrence of the solution in a certain region of  the invariant subspace associated with slow modes displays some extended predictability, 
as reflected by the oscillatory behavior of the error for the atmospheric variables at long lead times.

\end{abstract}


\end{frontmatter}

\section{Introduction and motivation}
\label{sec:intro}

The variability at annual, interannual and decadal time scales of the coupled ocean--atmosphere system is currently a central
concern in improving extended-range weather and climate forecasts. The oceans' long-term variability and
their interaction with the atmosphere has been extensively explored in the Tropical Pacific, due to the  
climatological importance of the El Ni\~no--Southern Oscillation (ENSO) phenomenon (e.g. \cite{Duan2013, Phil90, Stuecker2013}). The  
ocean and the atmosphere also interact in the mid-latitudes through both wind stress and buoyancy fluxes  \cite{Dijkstra2005, Vallis2006}), although the 
impact of this interaction on the long-term variability of the atmosphere is still controversial \citep{Brachet2012, Kushnir2002}.

On physical grounds, interactions between the two components of the coupled climate system in mid-latitudes are obviously
essential to its functioning on multiple time scales.  The main direction of the coupling, however, is a matter of debate:
Is the slower ocean responding to the wind stress forcing in an essentially passive way \cite{Frank77} --- i.e., is its
feedback to the atmosphere too weak to qualitatively modify the dynamics of a stand-alone atmosphere
--- while, at the same time, playing the role of a heat bath that provides boundary forcing for the atmosphere \cite{Roberts2000}?
Or is the ocean playing a more active role in atmospheric dynamics  \cite{FGS'04, FGS'07, Jiang1995}? 
\citet{Marshall2001}, for instance, discuss these questions in detail. 

From a dynamical point of view, one possible answer to these questions translates into a search for
the presence of coupled modes between the ocean and the atmosphere, such as found for instance in observational data \cite{Czaja2002}.
In the context of coupled global climate models, the answer, however, differs from one  
investigation to another or from a modeling setup to another; this answer depends, to a large extent, on whether the forcing is generated by deterministic or stochastic, linear or nonlinear processes 
\cite{Delworth2000, Kravtsov2008, Liu2012}. 

Part of our motivation is that a low-frequency nonlinear oscillation has been found in intermediate-order, coupled nonlinear ocean-atmosphere models. This coupled mode operates by triggering the displacement of the atmospheric jet position \cite{Kravtsov2007, VanderAvoird2002}. 
The physical origin of this coupled mode is, however, not fully understood as yet, 
nor is its existence generally agreed upon.

To understand the qualitative behavior of the coupled ocean--atmosphere dynamics, 
several low-order models have been  developed. 
Such models allow one to isolate the essential processes believed to be at play in a specific problem at hand, by using as building blocks only 
the minimal ingredients describing the dynamics.

This approach --- originating in the works of Saltzman \cite{Saltzman1962} and Lorenz \cite{Lorenz1963} and, from then on, in the development and applications of
nonlinear dynamics to the environmental sciences --- attempts to reduce complicated dynamics to its essential
features. It has been quite successful, so far, in increasing our understanding of the dynamics of the ocean alone, in particular the multimodality of the thermohaline circulation 
\cite{Dijkstra2005, Stommel1961}, as well as the development and variability of the mid-latitude oceanic gyres
\cite{Jiang1995, Pierini2012, Simonnet2005, Veronis1963}.

To the best of our knowledge, it is Lorenz \cite{Lorenz1984} who applied this approach first to the coupled system 
and developed a pseudo-spectral, low-order model based on the primitive equations for the atmosphere, coupled to an ocean heat bath. Vannitsem \cite{Vannitsem2011} used this model 
to evaluate the impact of climate changes on model error biases.
Nese and Dutton \cite{Nese1993} extended the model by adding an ocean dynamics 
similar to the one proposed by \citet{Veronis1963} and based on four dominant ocean modes. These authors found that accounting for the heat transport 
helps increase the coupled model's predictability. 

Other minimal-order coupled models \cite{Roebber1995, VanVeen2003} allowed for the possibility of the ocean's developing a thermohaline circulation. In the latter work,
\citet{VanVeen2003} performed a bifurcation analysis and showed the active role of the ocean in setting up the dynamics when close to the bifurcation points of 
the atmospheric model.  While \citet{Deremble2012} did not consider full coupling, they did point out some of the similarities between the bifurcation trees 
of the atmospheric and oceanic dynamics in a mid-latitude setting.

More recently, Vannitsem and colleagues \cite{Vannitsem2013, Vannitsem2014} have developed two coupled model versions based on a quasi-geostrophic atmospheric model 
proposed by \citet{Charney1980} and further studied in \cite{Reinhold1982, Reinhold1985}, and on the ocean model of \citet{Pierini2012}. The coupling in both of 
these versions was based solely on a mechanical transfer of momentum via wind friction. In particular,
these authors showed that their coupled model displays 
decadal variability within the ocean, similar to that found in idealized, intermediate \cite{Jiang1995, Speich1995} and 
low-order models \cite{Broer2011, Nadiga2001, Simonnet2005}. Furthermore, the investigation of this coupled model's stability properties 
showed that the momentum transfer coupling between the two components contributes to an increase of the flow's instability,  
in terms of both the magnitude of the positive  Lyapunov exponents and their number.  

This model \cite{Vannitsem2013, Vannitsem2014} is, however, missing an important ingredient of the coupled dynamics, namely 
the energy balance between the ocean and the atmosphere. The present work proposes a new model version,  
in which the thermal forcing affecting only the atmosphere is replaced by an energy balance scheme
\cite{Barsugli1998, Chen1996, Deremble2012}. It will allow us to disentangle the respective roles of the heat and radiative fluxes through the ocean surface vs. the transport of heat within the ocean 
in affecting the coupled system and its predictability.                      

The model equations are described in Section \ref{sec:dyn} 
, for the dynamics \cite{Vannitsem2013, Vannitsem2014} and thermodynamics, respectively; they are reduced to a low-order system in Section~\ref{sec:low-dim}. 
In Section~\ref{sec:results}, a bifurcation analysis of the basic solutions is first performed (Section~\ref{ssec:bif'n}); it reveals the presence of low-frequency variability (LFV) in the form of a  
set of long-periodic, attracting orbits that couple the dynamical modes of the ocean and the atmosphere. 
The model dynamics is further explored through the analysis of the climatological properties
of the solutions in Section~\ref{ssec:clim}, while  the dependence of the decadal-scale orbits on model parameters and their predictability are 
studied in Sections~\ref{ssec:slow_mfd} and \ref{ssec:Lyap}, respectively. A summary of the results and future prospects are then provided in Section~\ref{sec:concl}.

\section{The model equations}
\label{sec:dyn}

\subsection{Equations of motion for the atmosphere}
\label{ssec:atmos}

The atmospheric model is based on the vorticity equations of a two-layer, quasi-geostrophic
flow defined on a $\beta$-plane \cite{Pedlosky87, Vallis2006}. The equations in pressure coordinates are
\begin{eqnarray}
\frac{\partial}{\partial t} \left( \nabla^2 \psi^1_{\rm a} \right) + J(\psi^1_{\rm a}, \nabla^2 \psi^1_{\rm a}) + \beta \frac{\partial \psi^1_{\rm a}}{\partial x}
& = & -k'_d \nabla^2 (\psi^1_{\rm a}-\psi^3_{\rm a}) + \frac{f_0}{\Delta p} \omega, \nonumber \\
\frac{\partial}{\partial t} \left( \nabla^2 \psi^3_{\rm a} \right) + J(\psi^3_{\rm a}, \nabla^2 \psi^3_{\rm a}) + \beta \frac{\partial \psi^3_{\rm a}}{\partial x}
& = & +k'_d \nabla^2 (\psi^1_{\rm a}-\psi^3_{\rm a}) - \frac{f_0}{\Delta p}  \omega \nonumber \\  
& & - k_d \nabla^2 (\psi^3_{\rm a}-\psi_{\rm o}); 
\label{eq:atmos}
\end{eqnarray}
here $\psi^1_{\rm a}$ and $\psi^3_{\rm a}$ 
are the streamfunction fields at 250 and 750 hPa, respectively, and $\omega = dp/dt$ is the vertical velocity, 
$f_0$ is the Coriolis parameter at latitude $\phi_0$, with $\beta = df/dy$ its meridional gradient there. 

The coefficients $k_d$ and $k'_d$
multiply the surface friction term and the internal friction between the layers, respectively, while $\Delta p = 500$ hPa is the pressure difference between 
the two atmospheric layers.  An additional term has been introduced in this system in order to account for the presence of a surface boundary velocity
of the oceanic flow defined by $\psi_{\rm o}$ (see the next subsection). This would correspond to the Ekman pumping on a moving surface and is the mechanical
contribution of the interaction between the ocean and the atmosphere.

Equation \eqref{eq:atmos} has been nondimensionalized, as discussed in \ref{ssec:nondim}.

\subsection{Equations of motion for the ocean}
\label{ssec:ocean}

The ocean model is based on the reduced-gravity, quasi-geostrophic shallow-water model on a $\beta$-plane (e.g., \cite{GFS2002, Pierini2012, Vallis2006}):
\begin{equation}\label{eq:QGSW}
\frac{\partial}{\partial t} \left( \nabla^2 \psi_{\rm o} - \frac{\psi_{\rm o}}{L_R^2} \right) + J(\psi_{\rm o}, \nabla^2 \psi_{\rm o}) + \beta \frac{\partial \psi_{\rm o}}{\partial x}
= -r \nabla^2 \psi_{\rm o} + \frac{ {\mathrm{curl}}_z \tau}{\rho h}.
\end{equation}
Here $\psi_{\rm o}$ is the streamfunction in the model ocean's upper, active layer, which has density $\rho$, depth $h$, and lies over 
a quiescent deep layer with density $\rho_{\infty}$; 
$g'=g (\rho_{\infty} - \rho)/\rho$ is referred to as the reduced gravity felt by the fluid in the active layer, and
$L_R = \sqrt{g'h}/f_0$ is the reduced Rossby deformation radius. 
The friction coefficient at the bottom of the 
active layer is $r$, and ${\mathrm{curl}}_z \tau$ is the vertical component of the curl of the wind stress. 

Usually, in low-order ocean modeling, this curl is 
prescribed as an idealized profile, zero along a latitude $\phi_0$ placed at or near 45$^\circ$N, and
antisymmetric about this latitude \cite{Jiang1995, Simonnet2005}. In the present work, this forcing is provided 
by the wind stress generated by the atmospheric component of the coupled system. 

Assuming that the wind stress is given by $(\tau_x, \tau_y)=C (u-U,v-V)$ --- where $(u = -\partial \psi^3_{\rm a}/\partial y, v = \partial \psi^3_{\rm a}/\partial x)$ 
are the horizontal components of the geostrophic wind, and $(U, V)$ the corresponding components of the geostrophic currents
in the ocean --- one gets
\begin{equation}
\frac{\mathrm{curl}_z \tau}{\rho h} = \frac{C}{\rho h} \nabla^2 (\psi^3_{\rm a}-\psi_{\rm o}). 
\label{eq:stress}
\end{equation}
Here the wind stress is proportional to the relative velocity between the flow in the ocean's upper layer and the wind in the lower atmospheric layer. The drag coefficient
$d = C/(\rho h)$ characterizes the strength of the mechanical coupling between the ocean and the atmosphere and will be a 
key bifurcation parameter in the present work. 

Equation \eqref{eq:QGSW} is also made nondimensional, cf.  \ref{ssec:nondim}.

\subsection{Ocean temperature equation}
\label{ssec:ocean_temps}

We assume here that temperature is a passive scalar transported by the  
ocean currents, but the oceanic temperature field
displays strong interactions with the atmospheric temperature through radiative and heat
exchanges \cite{Barsugli1998, Deremble2012}.
Under these assumptions, the evolution equation for the ocean temperature is

\begin{equation}\label{eq:heat_oc}
\gamma_{\rm o} ( \frac{\partial T_{\rm o}}{\partial t} + J(\psi_{\rm o}, T_{\rm o})) = -\lambda (T_{\rm o}-T_{\rm a}) + E_{{\rm o},R}.
\end{equation}
with
\begin{equation}\label{eq:fluxes_oc}
E_{{\rm o},R} = -\sigma_B T_{\rm o}^4 + \epsilon_{\rm a} \sigma_B T_{\rm a}^4 + R_{\rm o}.
\end{equation}

In Eqs.~\eqref{eq:heat_oc} and \eqref{eq:fluxes_oc} above, $E_{{\rm o},R}$ is the net radiative flux at the ocean surface. This flux contains three terms: 
the shortwave radiation $R_{\rm o}$ entering the ocean;  the outgoing longwave radiation flux $-\sigma_B T_{\rm o}^4$; and 
the longwave radiation flux $\epsilon_{\rm a} \sigma_B T_{\rm a}^4$ re-emitted to the ocean; here $\epsilon_{\rm a}$ is the emissivity of the atmosphere 
and $\sigma_B$ is the Stefan-Boltzmann constant.
The parameter  $\gamma_{\rm o}$ is the heat capacity of the ocean, and $\lambda$ is the
heat transfer coefficient between the ocean and the atmosphere
that combines both the latent and sensible heat fluxes.  We assume that this combined heat transfer is
proportional to the temperature difference between the atmosphere and the ocean.
The typical values of the above parameters are provided in  Table~\ref{tab:dim_param}.

\subsection{Atmospheric temperature equation}
\label{ssec:atmos_temps}

Let us now consider the equation for the temperature of the baroclinic atmosphere presented in Section~\ref{ssec:atmos}.
As in the ocean, a radiative and heat flux scheme is incorporated reflecting the exchanges
of energy between the ocean, the atmosphere and outer space \cite{Barsugli1998}:

\begin{equation}\label{eq:heat_atm}
\gamma_{\rm a} ( \frac{\partial T_{\rm a}}{\partial t} + J(\psi_{\rm a}, T_{\rm a}) -\sigma \omega \frac{p}{R}) = -\lambda (T_{\rm a}-T_{\rm o}) + E_{{\rm a},R}
\end{equation}
with
\begin{equation}\label{eq:fluxes_atm}
E_{{\rm a},R} = \epsilon_{\rm a} \sigma_B T_{\rm o}^4 - 2 \epsilon_{\rm a} \sigma_B T_{\rm a}^4 + R_{\rm a}.
\end{equation}

In Eq. \ref{eq:heat_atm}, $\psi_{\rm a}=(\psi^1_{\rm a}+\psi^3_{\rm a})/2$ is the atmospheric barotropic streamfunction, $E_{{\rm a},R}$ is the net 
radiative flux in the atmosphere,
$R$ the gas constant, $\omega$ the vertical velocity in pressure coordinates, and 
$$\sigma = - \frac{R}{p} \Big(\frac{\partial T_{\rm a}}{\partial p}- \frac{1}{\rho_{\rm a} c_p}\Big) $$
is the static stability, with $p$ the
pressure, $\rho_{\rm a}$ the air density, and $c_p$ the specific heat at constant pressure; here $\sigma$ is
taken to be constant. Note also that, thanks to the hydrostatic relation in pressure coordinates and to the
ideal gas relation $p=\rho_{\rm a} R T_{\rm a}$, the atmospheric temperature $T_{\rm a}$ can be expressed as
$T_{\rm a} = - (p/R) f_0 (\partial \psi_{\rm a}/\partial p)$.

As for the radiative flux in the ocean, in Eqs.~\eqref{eq:heat_oc} and \eqref{eq:fluxes_oc}, the
net radiative flux $E_{{\rm a},R}$ within the atmosphere is composed of three terms: the ingoing flux 
$\epsilon_{\rm a} \sigma_B T_{\rm o}^4$ of radiative energy effectively absorbed; 
the outgoing flux $- 2 \epsilon_{\rm a} \sigma_B T_{\rm a}^4$ re-emitted to the ocean and to space; and the shortwave radiative flux $R_{\rm a}$ absorbed 
directly by the atmosphere. 
Note that, in writing Eqs. \eqref{eq:fluxes_oc} and \eqref{eq:fluxes_atm}, we assume that the rates
of radiative emission and absorption are equal, i.e that emissivity is equal to absorptivity. 
This assumption is strictly valid only at thermodynamic equilibrium, but it 
can be safely applied to systems in local thermodynamic equilibrium, like the lower 
atmosphere, in which molecular collisional processes dominate the radiative processes \cite{Houghton1986}. 

This expression for $T_{\rm a}$ can then be used to combine Eqs.~\eqref{eq:heat_atm} and \eqref{eq:atmos},  
as done when deducing the quasi-geostrophic potential
vorticity equation \cite{Vallis2006}.

\subsection{Low-order model formulation}
\label{sec:low-dim}

In order to build a low-order model version, we follow
the truncated Fourier expansion approach in \cite{Charney1980, Jiang1995, Lorenz1963, Pierini2012, Reinhold1982, Saltzman1962, Simonnet2005, Vannitsem2014, Veronis1963}, i.e., the model fields are 
developed in a series of basis functions and truncated at a
minimal number of modes that still captures key features of the observed behavior. Both linear and nonlinear terms in the equations of motion are then projected onto the
phase subspace spanned by the modes retained, by using an appropriate scalar product. 

In the thermodynamic equations introduced in Sections  \ref{ssec:ocean_temps} and \ref{ssec:atmos_temps}, quartic terms appear in the radiative fluxes. 
To overcome this problem, we will take advantage of the small amplitude of temperature anomalies, as compared with the global mean, in order to linearize these 
terms. The details of this linearization are described in \ref{ssec:linearize}.

For the model's closed ocean basin, we use only sine functions, in order to avoid fluxes through the boundaries. For the atmosphere, no-flux boundaries are 
assumed in the meridional direction and periodicity in the longitudinal direction. Hence 
both sine and cosine functions are allowed
in the longitudinal direction, while we use  
cosine modes only in the meridional direction,
as discussed for instance in \cite{Reinhold1982}.
The radiative fluxes are taken as proportional to a cosine function of latitude.

To obtain the coupled-model equations for the atmosphere, we combine the truncated temperature equations~\eqref{temp} 
of \ref{ssec:appendix} with the equations for the atmospheric streamfunction 
perturbation that result from the projection of the dynamical equations \eqref{eq:atmos} onto the modes in the set~\eqref{eq:atmos_set}.
As in \cite{Reinhold1982}, one obtains a set of $n_{\rm a} = 20$ ordinary differential equations (ODEs) for the atmosphere. 

Similarly, combining the projection of the streamfunction field for the ocean onto the modes of \eqref{eq:ocean_set} with
the temperature field of (\ref{eq:temp_set_oc_de_1}) leads to a set of $n_{\rm o} = 16$
ODEs, cf. \ref{ssec:appendix}. The full coupled model is thus based on a set of $n_{\rm a} + n_{\rm o} = 36$ ODEs.
The time-dependent solutions of this system are obtained by numerical integration, using 
a second-order Heun method with a fixed time step $\Delta t = 0.01$ time units. Several
higher-accuracy methods have been tested but without affecting substantially the results reported herein. 

In our analysis of the dynamics of the coupled system, we will mostly focus on three
parameters, given in dimensional units, in order to ascertain 
the typical values that give rise to remarkable behavior. 
These three parameters are $C_{\rm o}$ [Wm$^{-2}$], $d$~[s$^{-1}]$ and $\lambda$ [Wm$^{-2}$K$^{-1}$]:
they correspond to the meridional variation in the radiative input from the Sun; the strength of the coupling between the ocean and the atmosphere, as an inverse of a response
time scale; and the intensity of the heat fluxes, respectively.

\section{Dynamics of the coupled model}
\label{sec:results}

\subsection{Bifurcation diagram}
\label{ssec:bif'n}

We start our analysis of the coupled model's dynamics by constructing its bifurcation diagram, 
cf. \cite{Dijkstra2005, GhCh87, SDG'09} and references therein. The AUTO software \cite{Doedel2007}
is used to follow solution branches by pseudo-arclength continuation and
detect local bifurcations.  The bifurcations will be explored in the two-dimensional parameter space of $C_{\rm o}$
and $d$, with $C_{\rm a} = C_{\rm o}/4$ and $\lambda~=~20$~Wm$^{-2}$K$^{-1}$ fixed.

In constructing the model's bifurcation diagram, we first fix the mechanical friction coefficient between the atmosphere and the ocean at a 
value of $d =10^{-8}$ s$^{-1}$, and vary the parameter  $C_{\rm o}$ that scales the energy absorption by the ocean 
between 0 and 400.  The $L^2$ norm
plotted in Fig.~\ref{fig:Hopf}a summarizes the successive bifurcations of the solutions. We define the $L^2$-norm $\|\cdot\|_2$ of a solution by 
\begin{center}
    \begin{tabular}{ccl}
      $\displaystyle\| {\bf x} \|_2 \, = \, \left( \sum_{i=1}^n \, x_i^2 \right)^{1/2}$ &  & for a fixed point ${\bf x}=(x_1,x_2,\ldots,x_n)$, and \\
      $\displaystyle\| {\bf x} \|_2 = \left(\frac{1}{T} \, \int_0^T {\bf x}(\tau)^2 \, d\tau \right)^{1/2}$ & & for a periodic orbit ${\bf x}(\tau)$ with ,
    \end{tabular}
\end{center}
where $T$ is the period.

When  $C_{\rm o} = 0,$ 
the system has only one fixed point, which remains stable as $C_{\rm o}$ increases, until a first Hopf bifurcation occurs at  
$C_{\rm o} \simeq 194.5$ Wm$^{-2}$ (Fig.~\ref{fig:Hopf}a); it is clearly subcritical.

The 
periodic orbits along this branch have very long periods, 
of roughly 21 years. This ``slow'' branch encounters a fold
at $C_{\rm o} \simeq 188.5$ Wm$^{-2}$ and, from there on, it stabilizes and 
extends forward as the parameter increases further. Such a fold is often associated with a ``global'' Hopf bifurcation, in the terminology of \cite{GhTav83}; see also \cite[Ch.~11]{GhCh87} and references therein. The stable periodic branch 
loses its stability again at $C_{\rm o} \simeq 271$ Wm$^{-2}$, according to the 
values of the Lyapunov exponents computed in Section~\ref{ssec:Lyap}.

Interestingly, the long-periodic branch is no longer present in Fig.~\ref{fig:Hopf}b, i.e.  at $d=0$. 
This essential difference between panels (a) and (b) of Fig.~\ref{fig:Hopf} indicates that the frictional coupling between ocean and atmosphere
is at the origin of the development of the long-periodic oscillation.  In other words, wind friction is, in the
present coupled model, the main physical process triggering the LFV development within {\it both} the atmosphere and the ocean.

After the first Hopf bifurcation, the fixed-point branch 
undergoes another Hopf bifurcation, at $C_{\rm o} \simeq 258$ Wm$^{-2}$. The 
periodic orbits that emanate from this bifurcation have short periods, of 
10--15 days, and are unstable. According to the AUTO software, this ``fast'' branch possesses many torus bifurcations, also referred to as secondary Hopf bifurcations. Indeed, depending on the value of $C_{\rm o}$, various stable quasi-periodic solutions are found in the neighborhood of this branch: these solutions are characterized by the coexistence of one long period, of about 22 years, and a short one, of about 5 to 20 days. Concomitantly, 
these solutions are associated with a weak transport, whose intensity is much smaller 
than the transport associated with the slow branch.

As shown in Fig.~\ref{fig:Hopf}b, the fast branch is already present in the bifurcation diagram when the ocean and the atmosphere are
not coupled mechanically, i.e. at $d=0$. For this value of $d$, the fast branch is stable until a torus bifurcation occurs at $C_{\rm o} = 272$ Wm$^{-2}$. The  
quasi-periodic solutions found after this bifurcation do not possess the above-mentioned long periodicity, only 
the short ones. These 
findings indicate the purely atmospheric origin of the latter, short-periodic oscillations. Indeed, the oceanic transport 
vanishes for the quasi-periodic solutions near this secondary Hopf bifurcation, $\psi_{\rm o} \equiv 0$.
Interestingly, along the fast branch and for the tori found nearby, all the atmospheric modes and 
a few of the ocean temperature modes ---  
namely $T_{{\rm o},i}$, with $i = 1,3,5$ --- 
are oscillating on  the fast time scale, while the other modes remain stationary.

Oscillatory atmospheric modes having periodicities that are longer than the typical life time of extratropical storms but shorter than a season, i.e. 
roughly of $10-100$ days, are called {\em intraseasonal} and have been described and discussed at some length in the dynamic-meteorology literature, especially 
in the Northern Hemisphere extratropics \cite{GhCh87, GhilRob'02}. Their impact on similarly fast oceanic modes in this coupled model is interesting but not entirely surprising.

The loci of the two Hopf bifurcations that lead to the slow and the fast 
periodicities intersect in a Hopf--Hopf bifurcation, as seen in the regime diagram of Fig.~\ref{fig:regime}. The locus associated with the slow branch finally  terminates 
in a Bogdanov-Takens bifurcation.  A locus of torus bifurcations emanates from the Hopf--Hopf bifurcation and accounts for the bifurcations of this type found along the  
fast branch in Fig.~\ref{fig:Hopf}a.
Another locus of torus bifurcations accounts for the bifurcation that is already present when $d=0$; see discussion of Fig.~\ref{fig:Hopf}b above.
The locus of the fold of the long-periodic orbit is depicted in red in Fig.~\ref{fig:regime} and we see that it merges with the curve of the Hopf bifurcation of the long-periodic orbit in a generalized Hopf bifurcation 
at $C_{\rm o} \simeq 170$ Wm$^{-2}$.

This bifurcation diagram provides only a partial view of our coupled model's rich dynamics, 
due to two factors: first, several additional parameters may trigger 
further qualitative changes in system behavior, like the heat flux parameter $\lambda$; and second, other 
bifurcation types, like global bifurcations or blue-sky catastrophes, 
might be present \citep{Kuznetsov1998, SDG'09}.  The rich behavior  presented in Sections~\ref{ssec:clim} and \ref{ssec:Lyap} below may 
arise, at least in part, from the presence of such nonlocal bifurcations.

Nevertheless, the local bifurcation diagram already reflects the wide variety of solutions generated by this
relatively simple model, and in particular the presence of a long-periodic solution associated with the coupling of the atmosphere with the ocean. We must emphasize 
that this bidecadal periodicity is not only a property of the ocean but also of the dominant
dynamical modes of the atmosphere, as illustrated in Fig.~\ref{fig:LCs}.  In this 
figure, we plot a three-dimensional (3-D) projection of the long-periodic orbits 
onto the subspace spanned by the modes $(\psi_{{\rm a},1}, \psi_{{\rm o},2}, T_{{\rm o},2})$, for $d = 10^{-8} s^{-1}$ and for
several $C_{\rm o}$-values. 

All the periodic orbits plotted in Fig.~\ref{fig:LCs} involve not just the three variables shown, but a total of $d_{\rm s} = 17$ variables --- namely $\psi_{{\rm o}, i}$ for $i=2,4,6,8$, $T_{{\rm o}, i}$ for $i=2,4,6$, 
$\psi_{{\rm a}, i}$ for $i = 1,5,6,9,10$, and 
$\theta_{{\rm a}, i}$ for $i = 1,5,6,9,10$ --- while the $d_{\rm f} = 19$ other variables are equal to 0. If the latter variables are 
set to 0 initially, they remain equal to 0 as the flow evolves. Hence the subspace of the 17 variables listed above is invariant under the phase space flow 
induced by the 36 ODEs that govern our coupled model. Moreover, all the orbits we computed within this subspace were dominated by slow motions; 
hence the use of the subscript `s' for ``slow'' and of `f' for ``fast'' with respect to the 19 other variables.  

The orbit shown for $C_{\rm o} = 270$ Wm$^{-2}$ (blue curve in Fig.~\ref{fig:LCs}) is
not only periodic but also attracting, i.e., it is a genuine, stable limit cycle. 
The other periodic orbits plotted in the figure, across the range of parameters $280 \le C_{\rm o} \le 320$ Wm$^{-2}$, while unstable for arbitrary perturbations, are actually stable to perturbations within the slow subspace,
even for very large values of $C_{\rm o}$. 

Perturbations introduced in the
complementary 19-dimensional subspace --- for values of $C_{\rm o}$ larger than about $271$~Wm$^{-2}$  --- do not die out, but saturate fairly quickly ; see Section~\ref{ssec:Lyap}, especially Fig.~\ref{fig:predict} there.
It is reasonable, therefore, to conjecture that the limit cycles in Fig.~\ref{fig:LCs} form the ``backbone'' of a strange attractor. 
The dimension $d_{\rm a} > 1$ of this attractor will be examined in Sections ~\ref{ssec:slow_mfd} and \ref{ssec:Lyap} below, while the precise meaning 
we attach to the term {\em backbone} will be discussed in Section~\ref{sec:concl}.

We will see in fact later that the LFV associated with the long-periodic orbits in Fig.~\ref{fig:LCs} --- whether stable or not --- is still present 
when chaotic solutions develop beyond a certain range of parameter values ($C_{\rm o}, d$).  The decadal-scale limit cycles thus continue to organize
the coupled-system dynamics, well beyond this range, and the chaotic solutions still live close to them. 

\subsection{Climatological properties for large values of $C_{\rm o}$}
\label{ssec:clim}

Before discussing in  further detail the model dynamics 
around the long-periodic orbits shown in Fig. (\ref{fig:LCs}), let us now focus briefly  on the climatological properties of typical solutions generated by the system. 
The numerical code is the same second order code already mentioned in the previous section.

Figure~\ref{fig:mean} displays the mean fields for the ocean and the atmosphere for $C_{\rm o} = 300$ Wm$^{-2}$ and $d=10^{-8}$ s$^{-1}$.
The ocean is displaying in panel (c) the double-gyre dynamics that is well known from ocean models with prescribed, time-independent wind stress 
\citep{Dijkstra2005, Jiang1995, SDG'09}, but now in a genuinely coupled ocean--atmosphere model and thus including the temperature field in panel (d).  
This double gyre is transporting --- in the present, coupled model --- heat toward the pole, and it thus reduces the heat contrast affecting the atmosphere through its interaction with the ocean, as is the case in the observations \citep{Dijkstra2005} and in much more detailed models \citep[and references therein]{Lheveder2014}. 

In addition, a clear high- and low-pressure dipole is appearing in the atmospheric streamfunction and temperature fields of Figs.~\ref{fig:mean}(a,b), with a low-pressure  
center in higher latitudes, north of 45$^\circ$N, and
a high-pressure center further south. 
The atmospheric dipole is present whatever the 
value of $C_{\rm o}$, i.e., it is independent of 
the nature of the dynamics, whether stationary, periodic or chaotic. 
This feature is also present when $d= 0$, indicating that it is induced by the model's radiative scheme 
(not shown).

For all parameter values examined, a mean jet is forming between the two pressure centers. In the present, coupled model, the
two pressure centers, and hence the jet between the two, are strengthened by the localized heat transfer between the atmosphere and the ocean.  

The mean dipole and the jet between the two ``semi-permanent centers of action'' are reminiscent of 
similar features in the real atmosphere over the North Atlantic and North Pacific \citep[and references therein]{Lorenz1967}.
For instance, the Icelandic low and the Azores high 
control the position and intensity
of the mid-latitude jet affecting the Atlantic and Western Europe.

The intensity of the low- and high-pressure centers also depends strongly in our coupled model on the
degree of mechanical coupling between the ocean and the atmosphere, as shown in Fig.~ \ref{fig:couple_2}.
Furthermore, the increase of the coupling $d$ reduces the temperature contrast within the ocean, which in turn reduces the one within the atmosphere; compare
panels (b) and (d) of Figs.~\ref{fig:mean} and \ref{fig:couple_2}. We 
suspect, therefore, that the predictability of the system
will be highly affected by the ocean transport, since a drastic reduction of the meridional temperature 
gradient is experienced. We investigate this aspect in Section~\ref{ssec:Lyap}, in which Lyapunov exponents are computed.

\citet{FGS'04, FGS'07} have shown that the mid-latitude jet is much stronger yet in the presence of a sharp sea surface temperature (SST) front, like the Gulf Stream or 
the Kuroshio. Such a sharp front is not possible in our intermediate, low-resolution coupled model, nor in typical general circulation models (GCMs) used until recently 
in climate change studies. When such a front is introduced into a GCM, the stronger jets predicted in \citep{FGS'04, FGS'07} do appear \citep{Brachet2012, Minobe2008}.

\subsection{ Decadal-scale dynamics and its dependence on coupling}
\label{ssec:slow_mfd}

In order to better understand the 
changes in model dynamics that occur when the mechanical coupling $d$ is increased, we have represented in Fig.~\ref{fig:slow_mfd} the projection of the trajectories 
onto the 3-D subspace spanned by the modes $(\psi_{{\rm a},1}, \psi_{{\rm o},2}, T_{{\rm o},2})$, as in Fig.~\ref{fig:LCs}.
For small values of $d$, the solutions of the coupled model are well localized around small values of $\psi_{{\rm o},2}$; see red and green trajectories in 
panels (a) and (b). When $d$ is increased, the ocean dynamics becomes more vigorous, 
the transport of heat toward the pole increases, and the amplitude of the LFV in this subspace increases. 
This increase induces the high-amplitude cycles displayed in Figs.~\ref{fig:slow_mfd}a and b.

Along these large limit cycles, when the meridional temperature gradients are intense, i.e. when $T_{{\rm o},2}$ is large,
the transport is strong, in order to reduce the gradient. At the same time, the heat transfer
out of the ocean induces a high variability within the atmosphere. Once the amplitude of the temperature
gradient associated with $T_{{\rm o},2}$ decreases, the oceanic transport becomes less intense and the atmospheric dynamics more quiescent.

Another important and recently much-debated question (e.g., \citet{Delworth2000})
is the role played by the heat fluxes in the coupling
between the ocean and the atmosphere. 
The effect of $\lambda$ on the coupled model's LFV is shown in Fig.~\ref{fig:slow_lambda}, which shows the same type of trajectories as in Fig.~\ref{fig:slow_mfd}, 
but at different $\lambda$- rather than $d$-values.
The dynamics is drastically modified when increasing $\lambda$, with very chaotic trajectories at no heat flux (red) and a purely
periodic trajectory for the largest value of $\lambda$ in the figure. This  solution behavior suggests that the larger
the heat flux the stabler the dynamics.  This aspect of our results  will be further
explored in the next subsection, by computing the coupled model's Lyapunov exponents.

We saw that our coupled model successfully simulates semi-permanent highs and lows 
that have some similarity with the Azores High and the Icelandic Low, as well as multi-annual LFV. 
One might thus wonder whether it might generate spatial features similar to the North Atlantic Oscillation (NAO) and display
its decadal variability, as proposed by \citet{FGR'11} or by \citet{Kravtsov2007, Kravtsov2008}.

Figure~\ref{fig:LFV} shows the geopotential height difference between two points of the spatial domain --- located at ($\pi/n, \pi/4)$ and $\pi/n, 3 \pi/4)$, in the 
model's nondimensional coordinates --- within the atmosphere,  for different values 
of the radiative meridional gradient $C_{\rm o}$ and of the coupling parameter $d$. These points correspond roughly to high- and low-pressure centers 
in the model atmosphere. 

When the solutions  stay close to the previously discussed slow, long-periodic orbits, as they seem to do in Fig.~\ref{fig:LFV}a, a low-frequency oscillation is
clearly present for which the delay
between the dominant maxima in all three time series is roughly 20 years, with secondary maxima also present in-between. This succession of high-amplitude and low-amplitude
maxima is associated with a slight asymmetry in the system's attractor between negative and positive values of $\psi_{{\rm o},2}$, an asymmetry that
strongly affects the oceanic transport.   

The strong and fairly smooth LFV in Fig.~\ref{fig:LFV}a contrasts with the
considerably smaller-amplitude and much more irregular behavior apparent in
 Fig.~\ref{fig:LFV}b. The latter is essentially dominated by the atmospheric variability when the
meridional heating gradient $C_{\rm o}$ is large and the frictional coupling $d$ is small, as
is the case in panel (b). This difference further illustrates the importance of the underlying slow
 orbits found in the coupled system, around which the dynamics organizes for a certain range of parameters.

To disentangle the LFV present in these model simulations, we applied singular-spectrum analysis (SSA; \cite{Ghil_al2002, VG1989}) to the 
time series in Fig.~\ref{fig:LFV}. For the model run with $C_{\rm o}=350$ Wm$^{-2}$ and $d=5 \times 10^{-8}$s$^{-1}$, the SSA spectrum plotted in Fig. \ref{fig:SSA}a indicates one periodicity, 
with a period of 11 years, that is significant at the 95\% level; this periodicity reflects the succession
of consecutive maxima within all three runs in Fig.~\ref{fig:LFV}a. A longer period, though, of 
roughly 22 years --- which corresponds to the distance between the dominant maxima in the time series --- also agrees with the complete 
length of the limit cycles in Figs.~\ref{fig:LCs} and \ref{fig:slow_lambda}; this bidecadal period does not appear to be significant in the analysis so far.

In order to isolate this longer period, the reconstruction of the 11-year signal \cite{GhVa91, Ghil_al2002} was removed and the SSA
algorithm applied again on the residual time series, as shown in Fig. \ref{fig:SSA}b. A 21.9-year period is now significant
 at the 95\% level, as expected from the visual inspection of the time series. In addition, the third, fourth and fifth harmonic of this 22-year cycle --- at 7.3, 5.5 and 4.4 years --- are all significant at the same level, indicating the highly anharmonic nature of the bidecadal cycle, as apparent in Fig.~\ref{fig:LFV}a, too.

The discussion of these results is postponed to the paper's last section.

\subsection{ Stability and predictability of the coupled model} 
\label{ssec:Lyap}

\paragraph{Lyapunov exponents} 

In dynamical systems theory, the stability of a nonlinear system, like our coupled model, is usually quantified by evaluating its Lyapunov exponents. These exponents characterize the growth or decay of small differences in 
the system's initial data, and thus help generalize stability analysis from linear to nonlinear differential equations; they 
are evaluated in the so-called tangent space of a model trajectory \cite{Legras1996, Parker1989}.

Lyapunov exponents can be 
computed using orthogonal vectors $\{\e_i(t); i = 1, \ldots, n\}$ in the tangent space of a single model orbit,
\begin{equation}\label{eq:Lyap}
\sigma_i(t) = \frac{1}{\delta t} \ln \frac{|\e_i(t+\delta t)|}{|\e_i(t)|};
\end{equation}
here $\e_i(t)$ is the Lyapunov vector corresponding to the $i^{\rm th}$ Lyapunov exponent, in decreasing algebraic-value order \cite{Kalnay2003}, $n = 36$ is the dimension of the system, and $\delta t$ is the
 time step used in this computation. In the present section,  we compute our coupled model's Lyapunov exponents
and use them to explore the predictability properties of the flow as a function of the parameter values. To integrate the tangent linear system and rescale the Lyapunov vectors, we used the same
 second-order Heun scheme, with a time step of $\Delta t = 0.01$ time units, as for integrating the full nonlinear model in Sections~\ref{ssec:clim} and \ref{ssec:slow_mfd}.  

 Figures~\ref{fig:lead_Lyap}(a, b) display the effect of the radiative-forcing parameter $C_{\rm o}$ on the
three leading Lyapunov exponents, for $C_{\rm a}=C_{\rm o}/4$ and $C_{\rm a}=C_{\rm o}/3$, respectively.
 As seen in panel (a), model behavior experiences a transition from stationary --- for $C_{\rm o}=115, 125, 150$ and $175$ Wm$^{-2}$ --- through periodic, from $C_{\rm o}=190$ up to 
about $C_{\rm o}=270$ Wm$^{-2}$, and on to chaos, with two exponents becoming positive at almost the same value of $C_{\rm o}$.

For a larger shortwave radiative forcing within the atmosphere, $C_{\rm a}=C_{\rm o}/3$ (panel b), the transition 
occurs slightly earlier, at $C_{\rm o} \simeq 250$ rather than $C_{\rm o} \simeq 280$ Wm$^{-2}$, and
the amplitude of both positive exponents is somewhat larger. But the overall features of the transition are similar in the two panels. We will therefore
focus in the following on the case for which  $C_{\rm a}=C_{\rm o}/4$.

As mentioned in Section~\ref{ssec:bif'n}, we evaluated the stability of the slow branch --- green solid curve in  Fig. \ref{fig:Hopf}a 
for stable solutions and green dashed for unstable solutions ---  by computing the Lyapunov exponents. To do so, we followed the algorithm outlined above,
starting from initial
points close to the long-periodic orbit, for each set of parameter values explored. Stable periodic solutions, i.e. genuine limit cycles, correspond to
the first exponent in Fig. \ref{fig:lead_Lyap}a being zero, and they lie precisely along the solid green line displayed in Fig. \ref{fig:Hopf}a. 
The detailed mechanism of destabilization of this slow limit cycle is not clear yet and it is probably associated with a global bifurcation, whose analysis is 
beyond the scope of the present study.     

Figure~\ref{fig:Lyap_spectrum} shows the complete Lyapunov spectra for $C_{\rm o} = 350$ Wm$^{-2}$, in red for $d=10^{-8}$ s$^{-1}$ and in blue for $d = 5 \times 10^{-8}$ s$^{-1}$.
The spectra indicate that the solutions are chaotic for both values of $d$.  But clearly a drastic change occurs when one increases the mechanical coupling parameter $d$, 
namely a substantial drop in the positive Lyapunov exponents. This drop is consistent with the emergent
effect of slow limit cycles on their vicinity in phase space, as already discussed in 
connection with Figs.~\ref{fig:Hopf}c, \ref{fig:slow_mfd} and \ref{fig:slow_lambda} in the previous three subsections.

The key role of these slow solutions in the coupled model's dynamics is further illustrated in Fig.~\ref{fig:KS}, in which we plotted
the phase space distribution of the local Kolmogorov-Sinai entropy $h_{\rm KS}$, given by the sum of the positive local Lyapunov exponents ,
\begin{equation}
h_{\rm KS}(\x) = \sum_{i=1}^L \sigma_i(\x);
\end{equation}
here $\x$ is the state vector in the model's 36-dimensional phase space, and $L$ is the number of positive local Lyapunov values. Kolmogorov-Sinai entropy, also called metric entropy, is discussed in \citep{Sinai'59}, and local Lyapunov exponents in \citep{Abarbanel1992}. The local metric entropy $h_{\rm KS}(\x)$ gives a measure of the overall divergence of trajectories in a neighborhood of a point $\x$, although local Lyapunov exponents, unlike the global ones, are not invariant under a nonlinear change of variables.

Figure~\ref{fig:KS} clearly indicates that the neighborhood of the slow periodic solutions is quite stable at
the parameter values used in the figure. When the ocean transport is more intense and the meridional temperature gradient is larger than in the run used in Fig.~\ref{fig:KS}, 
the flow is much  more unstable, as the atmospheric dynamics is much more active (not shown).

\citet{Vannitsem1997} have investigated the properties of local Lyapunov exponents and of the corresponding local Kolmogorov-Sinai entropy in a
three-level quasi-geostrophic model with a fairly realistic climatology and variability for the extratropical atmosphere \citep{Kravtsov2009, Marshall1993}.
In that model --- often referred to as the QG3 atmospheric model \cite{Kondrashov2004} --- the values of the local exponents were highly dependent 
on the large-scale weather regimes \citep{Vannitsem1998, Vannitsem2001}. In the present analysis, a similarly strong dependence on large-scale flow features is also
evident, as indicated by the large variability of the  Kolmogorov-Sinai entropy when moving along the slow  solutions in Fig.~\ref{fig:KS}.

Figure~\ref{fig:lead_Lyap_d} illustrates the change in stability as a function of the coupling parameter $d$. For $C_{\rm o} = 300$~Wm$^{-2}$, the increase in $d$
first leads to a sharp drop in the leading Lyapunov exponents. Increasing $d$ further induces a transition to the periodic solution already illustrated in 
Section \ref{ssec:slow_mfd}. For $C_{\rm o}=350$~Wm$^{-2}$ the picture is slightly different, since there is no transition to a periodic solution for the set of values explored. The solution of the system is
 still chaotic, but it is living close to the slow periodic solutions that do exist when $d$ is large enough. 
 
Changes in the heat flux parameter $\lambda$ can also affect strongly the stability of coupled-model solutions. In Fig.~\ref{fig:LyaP_heat}, we plot the leading 
Lyapunov exponent $\sigma_1$ as a function of $\lambda$ for different combinations of the other two parameters, $C_{\rm o}$ and $d$.   
Clearly $\sigma_1$ decreases as a function of $\lambda$, except for $C_{\rm o} = 300$ Wm$^{-2}$ and $d = 10^{-8}$ s$^{-1}$. The decrease is sharpest for $C_{\rm o} = 350$ Wm$^{-2}$ and $d = 10^{-8}$ s$^{-1}$ (dotted blue curve) and weakest for $C_{\rm o} = 300$ Wm$^{-2}$ and $d = 3 \times 10^{-8}$ s$^{-1}$ (dotted green curve).
Overall, these results suggest that the heat fluxes between the ocean and the atmosphere have a stabilizing effect on the coupled model.

This stabilizing effect of the fluxes is also seen in examining the dimension of the attractor associated with the slow behavior in our coupled model. We consider the Kaplan-Yorke dimension $d_{\rm{KY}}$ \cite{KY1979}, often called the Lyapunov dimension; it is given by a simple functional of the Lyapunov exponents,
$$d_{\rm{KY}} = k^* + \frac{\sum_{k = 1}^{k^*} \lambda_k}{| \lambda_{k^*+1} |}, $$ 
where $k^*$ is the largest $k$ such that  $\sum_k \lambda_k > 0$. and, under fairly general circumstances, it equals the information dimension. Hence, $d_{\rm{KY}}$
is preferable to several other ways of estimating the dimensionality of attractors \cite[Ch.~2]{Tel2006}.

The results are very instructive, indeed, for the three orbits plotted in Fig.~\ref{fig:slow_lambda} and for two additional ones, also computed at $C_{\rm o} = 350$~Wm$^{-2}$ and $d = 6 \times 10^{-8}$~s$^{-1}$. For the slow limit cycle at $\lambda=100$ Wm$^{-2}$K$^{-1}$ (blue curve in the figure), one obviously has $d_{\rm{KY}} = 1.$ As $\lambda$ decreases, one gets: for $\lambda=50$ (not in the figure) $d_{\rm{KY}} \simeq 8.7$, for $\lambda=20$ (green curve) $d_{\rm{KY}} \simeq 14.6$,  for $\lambda=1.0$ (not in the figure) $d_{\rm{KY}} \simeq 20.4$, and for $\lambda=0$ (i.e., no heat flux at all; red curve in the figure) $d_{\rm{KY}} \simeq 21.5$. Visually, the green and red objects in Fig.~\ref{fig:slow_lambda} look successively stranger, while our Kaplan-Yorke dimension calculations indicate that the corresponding attractors, at $C_{\rm o} = 350$~Wm$^{-2}$ and $d = 6 \times 10^{-8}$~s$^{-1}$, have
$d_{\rm{KY}} < d_{\rm s} = 17$ for all but the two lowest $\lambda$-values, namely 1.0 and 0.

\paragraph{Predictability} Finally, the long-term predictability of the atmospheric component is explored when the slow variability is
starting to dominate coupled-model dynamics. The experiment consists in evaluating the growth of initial errors in forecasts that are issued
for $C_{\rm o} = 300$ Wm$^{-2}$ and $d = 5 \times 10^{-8}$ s$^{-1}$. 

The errors are introduced into all the $n = 36$ variables of the coupled model, but we are most interested in error evolution for the $n_{\rm a} = 20$ atmospheric modes: their root-mean-square growth is tracked by restricting the $L_2$-norm defined in Section~\ref{ssec:bif'n} to  these 20 modes
according to
$$\| \tilde \x^{(a)}(t) - \x'^{(a)}(t) \|_2 \, = \, \left( \sum_{i=1}^{n_{\rm a}} \, (x_i-x'_i)^2(t) \right)^{1/2};$$ 
here $\tilde \x^{(a)}(t)$ and $\x'^{(a)}(t)$ are the projection of the control and perturbed trajectories, respectively, onto the atmospheric variables.

An ensemble of 1000 realizations was obtained by using random initial errors sampled from a Gaussian distribution with mean 0 and variance $10^{-12}$,
in nondimensional units, around two specific locations on the coupled model's attractor. These two locations are indicated in Fig.~\ref{fig:KS} by the black filled circle and the green triangle, respectively, and 
the error growth curves are displayed in Fig.~\ref{fig:predict}.

For short times and for the first initial state (black curve) the error grows very rapidly up to a time of the order of a few
days. This contrasts with the short-time behavior of the error for the second initial state (green curve), which stays quite small and thus indicates
potential for an accurate extended forecast beyond several years. The difference between the short-time error growth in the two cases reflects, of course, the difference in
local stability along the attractor, as revealed by the Lyapunov exponents (not shown) and the Kolmogorov-Sinai entropy; see Fig.~\ref{fig:KS}.

For long lead times, the error evolution exhibits a large-amplitude cycle, with large errors when the system is in the high-atmospheric-variability region along the attractor  
and small errors when the system revisits the region of low atmospheric variability; see again Fig.~\ref{fig:KS}. As discussed in Section~\ref{ssec:slow_mfd}, 
these two regions coincide with the temperature gradient within the ocean.

The large up- and downswings in error evolution for long lead times in our coupled model stand in sharp contrast  with the
error evolution in certain stand-alone atmospheric models,  like Lorenz's
low-order model of a moist general circulation \citep{Lorenz1984, Nicolis1995} or the already mentioned QG3 model \citep{Marshall1993, Vannitsem1997}.
In these atmospheric models, the error growth is predominantly monotonic --- like in typical high-resolution numerical weather prediction models \cite{Kalnay2003} --- and saturates quickly 
at an asymptotic mean value after a typical time scale of the order of the inverse of the first Lyapunov exponent.  

The oscillatory nature of the error evolution in the coupled ocean--atmosphere model at hand is the by-product of the modulation of the atmospheric variability by the oceanic modes, 
in the model regime in which the destabilized slow limit cycle still affects substantially the chaotic variability that replaces the purely periodic flow in the model's phase space. This nearly periodic modulation 
of the error evolution --- which is clearly visible in both the black and the green curves in Fig.~\ref{fig:predict} --- implies, in turn, that the typical time scale of convergence of the probability density of the coupled
model's variables toward their asymptotic probability density must be much larger than the time scale associated with the dominant Lyapunov exponent.
Such a slow convergence --- if confirmed by modeling studies with higher resolution and additional physical processes --- would provide some hope for longer-term, decadal-scale forecasts of the
coupled ocean-atmosphere system.

\section{Concluding remarks}
\label{sec:concl}

We have presented a low-order model that incorporates several key physical ingredients of the coupled ocean-atmosphere system at mid-latitudes:  
baroclinic instability in a dry atmosphere, upper-ocean mass and heat transport in a shallow
layer, an energy balance scheme incorporating radiative and heat fluxes, and a mechanical coupling mechanism between the 
 atmosphere and the ocean. The latter coupling allows for the development of vigorous, albeit smooth oceanic gyres, cf. Figs.~\ref{fig:mean}--\ref{fig:couple_2}.
The thermodynamic portion of the model represents a significant addition to previous model versions \cite{Vannitsem2013, Vannitsem2014}. and is responsible for some of the most striking results reported herein.

This model contains $n = 36$ variables that represent the dominant modes of variability of the coupled system, with $n_{\rm a} = 20$ variables for the atmosphere and $n_{\rm o} = 16$ for the ocean. Given 
the relatively large number of variables and processes involved, our model contains several 

parameters that allow us to calibrate the intensity of the heat and momentum exchanges between the 
atmosphere and the ocean. The three most important ones are the wind stress coupling $d$ between the two fluid media,
the meridional variation of  gradient $C_{\rm o}$ of the radiative flux coming from the sun, and the strength $\lambda$ of the heat fluxes between the two.

Our  bifurcation analysis and numerical integrations 
have revealed a rich variety of solutions and a diversity of regimes of model behavior; see Fig.~\ref{fig:Hopf}.
In particular, we found a coupled ocean--atmosphere mode for large values of the friction coupling $d$ and of the radiative-flux gradient $C_{\rm o}$. This 
coupled mode has a decadal time scale and it appears as a stable limit cycle for a range of parameter values, e.g. $190 \le C_{\rm o} \le 270$~Wm$^{-2}$ at $d = 10^{-8}$~s$^{-1}$; see green solution branch in Fig.~\ref{fig:Hopf}a, the stable limit cycles in Fig.~\ref{fig:LCs}, as well as the purely periodic solutions plotted in Figs.~\ref{fig:slow_mfd} and \ref{fig:slow_lambda}.

Interestingly, the slow, decadal-scale solutions involve not just the oceanic variables, but some of the atmospheric ones as well, and correspond thus to true coupled modes. In fact, we found a subspace of dimension $d_{\rm s} = 17$ in the model's 36-dimensional phase space that is invariant under the model flow and involves only slow motions. Even when the limit cycles in Fig.~\ref{fig:LCs} lose their stability, they still seem to organize the phase space flow nearby into solutions that are dominated by slow motions; see the solutions that appear as small perturbations of a limit cycle in Figs.~\ref{fig:slow_mfd} and \ref{fig:slow_lambda}, as well as the upper panel of Fig.~\ref{fig:LFV}. All the solutions of this type are still dominated by a decadal and a bidecadal periodicity, as visually apparent in Fig.~\ref{fig:LFV}a and further documented by the SSA spectra in Fig.~\ref{fig:SSA}.

These results suggest that the slow limit cycles form the 
{\em backbone} of a strange attractor that is, at first, embedded in the slow, 17-dimensional subspace and only extends beyond it as the atmosphere gradually decouples from the ocean. In fact, already the results summarized in Fig.~\ref{fig:slow_lambda} show that, as the heat flux intensity $\lambda$ between the two fluid media decreases, the solutions become more agitated, with the fast part of the variance increasing in strength. 

We conjecture that this backbone corresponds to the least-unstable  
periodic orbit embedded in the chaotic attractor, as suggested by the discussion of Fig.~1 in \cite{Ghil_al2002}. The term is justified here 
since this orbit can be clearly identified in our model's singular 
spectrum as the main structure organizing its
dynamics and a dominant source of its LFV.

We used the Kolmogorov-Sinai entropy to visualize in Fig.~\ref{fig:KS} that the perturbations of the slow limit cycles still contain large portions of relatively quiescent phase space flow, while other portions become more agitated. Next, we used the Lyapunov dimension $d_{\rm{KY}}$ to track the evolution of the strange attractor we suspect is embedded in the slow subspace, as parameters change. This was done for the three orbits plotted in Fig.~\ref{fig:slow_lambda} and for several additional ones, computed at the same $C_{\rm o}$  and $d$ values.

For the slow limit cycle at $\lambda=100$ Wm$^{-2}$K$^{-1}$ (blue curve in the figure), one obviously gets $d_{\rm{KY}} = 1.$ As $\lambda$ decreases, $d_{\rm{KY}}$ increases, until it finally exceeds $d_{\rm s} = 17$ at $d_{\rm{KY}} \simeq 12.$
In fact, when  $\lambda=0$, i.e., when there is no heat flux at all between the two media, $d_{\rm{KY}} \simeq 21.5$. Visually, the green and red objects in Fig.~\ref{fig:slow_lambda} look successively stranger, in full agreement with our Lyapunov dimension calculations. 

We thus have substantial numerical evidence for the presence of a strange attractor in our coupled model, which is completely dominated by slow motions for sufficiently strong coupling between the atmosphere and ocean. As this coupling weakens, the dimension of the attractor increases and it supports more and more fast variance.

The slow variability associated with this attractor reflects the strong effects of the horizontal heat transport within the ocean on the atmospheric fields above.
This horizontal transport intensifies, to first order, when the meridional temperature gradient in the ocean is large, and 
it slows down when the gradient is small. Concomitantly, a large meridional gradient in the ocean imposes large meridional gradients of heat flux and longwave radiation in the atmosphere; the latter, in turn, induce an increase of
baroclinic instability and hence fast variability within the atmosphere, which enhance the heat transported toward the pole by the latter. 

In the coupled model studied herein, these feedbacks between atmospheric and oceanic transports are activated by the mechanical coupling between the ocean and the atmosphere, 
 via wind stress forcing --- which induces the development of gyres --- as well as by the vertical heat fluxes, which are proportional to $\lambda$. Both of these couplings, mechanical and thermal, are  
simplified here by modeling the temperature as a passive scalar. Our results on the effects of ocean--atmosphere coupling are plausible, given the fact that similar mechanisms have been found to act in GCMs \citep{Delworth2000, Lheveder2014}; still, it is well worth incorporating  a more active coupling between ocean dynamics and temperature in future versions of the present model.

The periodicities found in our model are decadal and bidecadal, cf. Figs.~\ref{fig:LFV}a and Fig.~\ref{fig:SSA}. The former periodicity agrees roughly with the variability peak in the coupled modes of the slightly more highly resolved atmosphere--ocean model of \citet{Kravtsov2007, Kravtsov2008}. In the latter, the near-decadal peak was associated with alternations between two atmospheric regimes characterized by the subtropical jet's latitude; this bimodality of the jet position and intensity does not seem to play a key role in the present model.

On the other hand, the NAO exhibits a peak in its variability at 7-8 years \cite[and references therein]{FGR'11}, rather than at 10--12 years, like here and in the coupled modes of \cite{Kravtsov2007, Kravtsov2008}. The mechanism that drives the 7-8-year peak in \cite{FGR'11} is the purely oceanic gyre mode of \cite{Dijkstra2005, Jiang1995, Speich1995} --- which arises even for time-independent wind stress forcing --- through this mode's impact on the atmospheric jet that forms above the SST front associated with the gyre mode, cf.~\cite{FGS'04, FGS'07, Brachet2012}.

To clarify further the nature of the decadal and multidecadal variability in the present coupled model --- and its relation, if any, with the LFV in and around the North Atlantic --- will require two things. First, the spatio-temporal aspects of the slow coupled mode will have to be examined using multi-channel 
SSA \cite{Ghil_al2002}, rather than just the single-channel version applied herein. The results of such an  
analysis can then be compared more closely with the observed spatio-temporal patterns.
Second,
additional modes, zonal as well as meridional, will have to be added within both the oceanic and the atmospheric component of the low-order model in order to clarify 
the robustness of the LFV to a more detailed spatial representation. 

The stability of the coupled model's phase space flow was measured by its leading Lyapunov exponents (Figs.~\ref{fig:lead_Lyap}, \ref{fig:lead_Lyap_d} and \ref{fig:LyaP_heat}), as well as by its local 
Kolmogorov-Sinai entropy (Fig.~\ref{fig:KS}). We found that these stability measures are highly dependent on the parameter values, in particular on the coupling parameter $d$. For small values of $d$,
the transport of heat within the ocean is relatively small and the coupled system's instability is essentially driven by the atmosphere; hence the Lyapunov exponents are quite large. 
Once $d$ attains a certain threshold --- which depends in turn  on the values of $C_o$ and of $\lambda$ --- the values of the Lyapunov exponents are drastically reduced, cf. Fig.~\ref{fig:lead_Lyap_d}, 
while the oceanic transport intensifies. 

The drop in the coupled model's short-term error growth is obviously associated with an increase in predictability; this increase, 
in turn, emphasizes the ocean's potential importance in modulating our ability to predict
both components of the coupled system at short lead times. 
Note that this oceanic effect was not present in the 
previous, 24-equation version of the model \cite{Vannitsem2014}.

The main difference between the two model versions is the absence of a temperature equation in the ocean and of an energy balance between the ocean and the atmosphere 
in  the earlier version: in \cite{Vannitsem2014}, 
the thermal energy source is provided through a restoring force toward climatological radiative equilibrium 
within the atmosphere. In  
the earlier version there was, therefore, no 
heat transport within the ocean, and in turn no thermal impact on the atmosphere. As 
the present analysis shows, this tranfer of energy plays a crucial role in the predictability of the coupled system.

At long lead times, the coupled model's predictability is essentially dominated by the role of the LFV (see again Fig.~\ref{fig:LFV}) and of the hypothetical slow attractor, cf. Fig.~\ref{fig:KS}. When 
the slow attractor is present long-lead
forecasts can be fairly accurate even for the atmospheric variables; see Fig.~\ref{fig:predict}.
It is, therefore, worth investigating whether such slow, coupled modes are indeed
found within the real coupled ocean-atmosphere system or, at least, within a much more detailed coupled model.

Clearly, our relatively low-order, 36-dimensional coupled model has produced a substantial number of interesting and stimulating results. Several of its
limitations have already been mentioned and we summarize the main ones here. 

First, we have been working with a closed, rectangular ocean basin, somewhat similar to those in \cite{Jiang1995, Speich1995}. 
The presence of oscillations between the cyclonic and anticyclonic gyres in the upper ocean for such a configuration is by now well known, although its robust presence in much more realistic intermediate 
models has also been demonstrated \cite[and references therein]{Dijkstra2005}.

Second, the emphasis on the wind-driven, double-gyre circulation neglects the importance of the buoyancy-driven, overturning circulation \cite{Dijkstra2005, Selten1999, Stommel1961} for decadal and multi-decadal climate variability \cite{Chang2015, Ghil'01}. Some aspects of both types of oceanic circulation were included, for instance, in \cite{Chen1996} and in \cite{Kravtsov2007, Kravtsov2008}. A related
caveat is the passive scalar character of the temperature, which does not feed back on the ocean dynamics.

Among the missing processes in our model formulation is the hydrological cycle, which
can affect substantially the atmospheric variability \cite{Lorenz1984}, as well as the heat fluxes between the atmosphere and the ocean \cite{Brachet2012, Chen1996}.
To address the robustness of some of the key results obtained herein --- including the presence of slow coupled modes and
of their impact on predictability --- will require removing some of these limitations and expanding further the methodology used in the present work.

\section{Acknowledgments}
 This work originated in discussions between MG and SV during the {\em Mathematics for the Fluid Earth} Programme held at the Isaac Newton Institute in Cambridge, UK, in Fall 2013. We have also benefited from discussions with C. Nicolis and H. Goosse. This work is partially supported by the Belgian Federal Science Policy Office under contracts SD/CA/04A and BR/12/A2/STOCHCLIM (SV, JD, LDC), by the U.S. National Science Foundation under grant OCE-1243175 (MG), and by the U.S. Office of Naval Research under MURI grant N00014-12-1-0911 (MG).

\appendix

\section{Nondimensional equations: Dynamics}
\label{ssec:nondim}

The equations \eqref{eq:atmos}--\eqref{eq:stress} are nondimensionalized by scaling horizontal distances by $L, (x'=x/L, y'=y/L)$, time $t$ by $f_0^{-1}$, the 
vertical velocity $\omega$ by $f_0 \Delta p$ and  the atmospheric and oceanic streamfunctions $\psi_{\rm a}$ and $\psi_{\rm o}$ by $L^2 f_0$. 
The parameters are also rescaled as $2 k = k_d/f_0, k' = k'_d/f_0, h'' = h'_d/f_0$,  $\beta' = \beta L/f_0$,
$\gamma= - L^2/L_R^2$, $r'=r/f_0$ and $\delta = d/f_0 = C/(\rho h f_0)$. 

The atmospheric and oceanic fields are expanded in
Fourier series over the domain $(0 \le x' \le 2\pi/n, 0 \le y' \le \pi)$, where
$n$ is the aspect ratio between the meridional and the zonal extents of the domain, 
$n = 2 L_y/L_x = 2 \pi L/ (2 \pi L/n)$. The values of the parameters just discussed above that are held fixed in the rest of the paper are
$\pi L = 5000$ km, $f_0=0.0001032$ s$^{-1}$, $2k=k'=0.04$, $n=1.5$, $r'=0.000969$, $\beta'=0.2498$, and $\gamma=-1741$; the value of the Coriolis parameter 
$f_0$ corresponds to the domain's axis of N--S symmetry at $\phi_0 = 45^\circ$N. 

\section{Linearization around a climatological temperature}
\label{ssec:linearize}

We assume here that 
\begin{subequations}\label{eq:lin}
\begin{align}
& T_{\rm a} = T_{\rm a,0} + \delta T_{\rm a}, \label{lin_atm} \\
& T_{\rm o} = T_{{\rm o},0} + \delta T_{\rm o}, \label{lin_oc}
\end{align}
\end{subequations}
where $T_{{\rm a},0}$ and $T_{{\rm o},0}$ are 
climatological reference temperatures, 
and $(\delta T_{\rm o}/T_{{\rm o},0})^2 + 
(\delta T_{\rm a}/T_{{\rm a},0})^2 \ll 1$.
Then
\begin{subequations}\label{eq:SB}
\begin{align}
& \epsilon_{\rm a} \sigma_B T_{\rm a}^4  =  \epsilon_{\rm a} \sigma_B ({T_{{\rm a},0}}^4 + 4 T_{{\rm a},0}^3 \delta T_{\rm a} + 6 T_{{\rm a},0}^2 \delta T_{\rm a}^2 
+ 4 T_{{\rm a},0} \delta T_{\rm a}^3 + \delta T_{\rm a}^4), \label{SB_atm} \\
& \sigma_B T_{\rm o}^4  =  \sigma_B ({T_{{\rm o},0}}^4 + 4 T_{{\rm o},0}^3 \delta T_{\rm o} + 6 T_{{\rm o},0}^2 \delta T_{\rm o}^2 
+ 4 T_{{\rm o},0} \delta T_{\rm o}^3 + \delta T_{\rm o}^4). \label{SB_oc}
\end{align}
\end{subequations}
We also let
\begin{subequations}\label{eq:rad}
\begin{align}
& R_{\rm a} = R_{{\rm a},0} + \delta R_{\rm a}, \label{rad_atm} \\
& R_{\rm o} = R_{{\rm o},0} + \delta R_{\rm o}, \label{rad_oc}
\end{align}
\end{subequations}
with $R_{{\rm o},0}$ and  $R_{{\rm a},0}$ the averaged shortwave radiative forcings.

Neglecting the higher-order terms in $\delta T$ in 
Eq.~\eqref{SB_oc} and
separating the reference
and the perturbation --- i.e., the zeroth-order terms in the expansion from the first-order ones ---  
leads to the following temperature equations for the ocean:
\begin{subequations}\label{eq:temp_oc_lin}
\begin{align}
\gamma_{\rm o} \frac{\partial T_{{\rm o},0}}{\partial t} = -\lambda (T_{{\rm o},0}-T_{{\rm a},0}) 
-\sigma_B T_{{\rm o},0}^4 + \epsilon_{\rm a} \sigma_B T_{{\rm a},0}^4 + R_{{\rm o},0},  \label{oc_mean} \\
\gamma_{\rm o} \Big( \frac{\partial \delta T_{\rm o}}{\partial t} + J(\psi_{\rm o}, \delta T_{\rm o})\Big) = -\lambda (\delta T_{\rm o}- \delta T_{\rm a}) \notag \\ 
-4 \sigma_B T_{{\rm o},0}^3 \delta T_{\rm o} + 4 \epsilon_{\rm a} \sigma_B T_{{\rm a},0}^3 \delta T_{\rm a}+ \delta R_{\rm o}. \label{oc_lin} 
\end{align}
\end{subequations}
Applying the same procedure 
to the atmospheric temperatures yields:
\begin{subequations}\label{eq:temp_atm_lin}
\begin{align}
\gamma_{\rm a} \frac{\partial T_{{\rm a},0}}{\partial t} = -\lambda (T_{{\rm a},0}-T_{{\rm o},0}) 
+ \epsilon_{\rm a} \sigma_B T_{{\rm o},0}^4 -2 \epsilon_{\rm a} \sigma_B T_{{\rm a},0}^4 + R_{{\rm a},0},   \label{atm_mean} \\
\gamma_{\rm a} \Big( \frac{\partial \delta T_{\rm a}}{\partial t} + J(\psi_{\rm a}, \delta T_{\rm a})-\sigma \omega \frac{p}{R}\Big) = -\lambda (\delta T_{\rm a}- \delta T_{\rm o}) \notag \\
+4 \epsilon_{\rm a} \sigma_B T_{{\rm o},0}^3 \delta T_{\rm o} - 8 \epsilon_{\rm a} \sigma_B T_{{\rm a},0}^3 \delta T_{\rm a}+ \delta R_{\rm a}. \label{atm_lin} 
\end{align}
\end{subequations}

It is interesting to note that the equations (\ref{oc_mean}, \ref{atm_mean})
for the climatological references are independent of the perturbations.
This implies that stationary solutions
can be readily found  by solving
\begin{subequations}\label{eq:stat}
\begin{align}
& -\lambda (T_{{\rm a},0}-T_{{\rm o},0}) + \epsilon_{\rm a} \sigma_B T_{{\rm o},0}^4 -2 \epsilon_{\rm a} \sigma_B T_{{\rm a},0}^4 + R_{{\rm a},0} = 0, \label{stat_atm} \\
& -\lambda (T_{{\rm o},0}-T_{{\rm a},0}) -\sigma_B T_{{\rm o},0}^4 + \epsilon_{\rm a} \sigma_B T_{{\rm a},0}^4 + R_{{\rm o},0} = 0. \label{stat_oc}
\end{align}
\end{subequations}
Summing the two equations
and substituting the expression obtained for $\epsilon_{\rm a} \sigma_B T_{{\rm a},0}^4$ into 
the first one leads to
\begin{equation}
\lambda (T_{{\rm a},0}-T_{{\rm o},0}) + 2  R_{{\rm o},0} + R_{{\rm a},0} - (2-\epsilon_{\rm a}) \sigma_B T_{{\rm o},0}^4 = 0,
\end{equation}
a quartic equation for the mean ocean temperature $T_{{\rm o},0}$. Solving this algebraic equation analytically by using Mathematica \citep{Wolfram2012} 
leads to a unique real and positive stationary solution $T_{{\rm o},0}$ for $R_{{\rm a},0}=R_{{\rm o},0}/4$, 
a typical ratio for the shortwave radiation absorbed by the atmosphere vs. that absorbed by the ocean.
Moreover, this stationary solution is stable for a wide range of values of $R_{{\rm o},0}$.

The stationary solutions so obtained, however, are not really close to the
observed values
for the coupled ocean--atmosphere system. 
In order to have our model operate in a more realistic 
domain of phase-parameter space, we
set $T_{{\rm a},0}=270$~K and $T_{{\rm o},0}=285$~K, as in \cite{Barsugli1998}.
This choice will only affect the values of the parameters
in front of the linearized radiative terms of 
Eqs.~\eqref{eq:temp_oc_lin} and \eqref{eq:temp_atm_lin}.

\section{Derivation of the truncated equations}
\label{ssec:appendix}

In this appendix, we describe how the low-order model is 
derived, by projecting the equations presented in section \ref{sec:dyn} onto a suitable orthogonal basis.

For the atmosphere, we keep the same set of modes as 
in \cite{Reinhold1982, Vannitsem2014},
\begin{eqnarray}\label{eq:atmos_set}
F_1 & = & \sqrt{2} \cos(\pi y/L_y) =  \sqrt{2} \cos(y'), \nonumber \\
F_2 & = & 2 \cos(2 \pi x/L_x ) \sin(\pi y/L_y) =  2 \cos(n x') \sin(y'), \nonumber \\ 
F_3 & = & 2  \sin(2 \pi x/L_x) \sin(\pi y/L_y) =  2  \sin(n x') \sin(y'), \nonumber \\ 
F_4 & = & \sqrt{2} \cos(2 \pi y/L_y) =  \sqrt{2} \cos(2y'), \nonumber \\
F_5 & = & 2  \cos(2 \pi x/L_x) \sin(2  \pi y/L_y) =  2  \cos(n x') \sin(2y'),  \nonumber \\ 
F_6 & = & 2 \sin(2 \pi x/L_x) \sin(2  \pi y/L_y) =  2 \sin(n x') \sin(2y'), \nonumber \\ 
F_7 & = & 2 \cos(4 \pi x/L_x) \sin( \pi y/L_y) =  2 \cos(2 n x') \sin(y'), \nonumber \\ 
F_8 & = & 2  \sin(4 \pi x/L_x) \sin( \pi y/L_y) =  2  \sin(2 n x') \sin(y'), \nonumber \\ 
F_9 & = & 2  \cos(4 \pi x/L_x) \sin(2  \pi y/L_y) =  2  \cos(2 n x') \sin(2y'), \nonumber \\ 
F_{10} & = & 2 \sin(4 \pi x/L_x) \sin(2  \pi y/L_y) = 2 \sin(2 n x') \sin(2y'),  
\end{eqnarray}
while the imposed radiative fluxes 
are given by $\delta R_{\rm a} = C_{\rm a} F_1(y)$ and  $\delta R_{\rm o} = C_{\rm o} F_1(y).$ 
One furthermore assumes that $C_{\rm a} = \alpha C_{\rm o}$, where $\alpha$ is
the fraction of shortwave radiation absorbed by the atmosphere. Note that two values of
this fraction will be used, $\alpha = 1/4$ and 1/3, 
both of which lie in a range that is typical for 
Earth's atmosphere.

For the ocean, we retain the following set of modes: 
\begin{eqnarray}\label{eq:ocean_set}
\phi_1 & = & 2 \sin(\pi x/L_x) \sin(\pi y/L_y) = 2 \sin(n x'/2) \sin(y'), \nonumber \\ 
\phi_2 & = &  2 \sin(\pi x/L_x) \sin(2 \pi y/L_y) = 2 \sin(n x'/2) \sin(2y'), \nonumber \\ 
\phi_3 & = & 2 \sin(2 \pi x/L_x) \sin(\pi y/L_y) = 2 \sin(n x') \sin(y'), \nonumber \\ 
\phi_4 & = & 2 \sin(2 \pi x/L_x) \sin(2 \pi y/L_y) =  2 \sin(n x') \sin(2y'), \nonumber \\ 
\phi_5 & = &  2 \sin(\pi x/L_x) \sin(3 \pi y/L_y) = 2 \sin(n x'/2) \sin(3y'), \nonumber \\ 
\phi_6 & = &  2 \sin(\pi x/L_x) \sin(4 \pi y/L_y)= 2 \sin(n x'/2) \sin(4y'), \nonumber \\ 
\phi_7 & = &  2  \sin(2 \pi x/L_x) \sin(3 \pi y/L_y) =  2  \sin(n x') \sin(3y'), \nonumber \\ 
\phi_8 & = & 2 \sin(2 \pi x/L_x) \sin(4 \pi y/L_y) =  2 \sin(n x') \sin(4y').
\end{eqnarray}

This set contains two additional modes in the latitudinal direction, as compared
to \citet{Veronis1963}; see also \citep{Jiang1995, Simonnet2005}.
This addition will allow us to get a more realistic temperature profile within the ocean.

We assume hereafter that one can project and truncate the atmospheric and oceanic streamfunction fields $\psi_{\rm a}(x, y)$ and $\psi_{\rm o}(x, y)$, as well as the corresponding temperature fields $\delta T_{\rm a}(x, y)$ and $\delta T_{\rm o}(x, y)$, and the vertical velocity $\omega(x, y)$ according to 

\begin{subequations}\label{eq:temp_set_atm}
\begin{align}
& \psi_{\rm a} = \sum_{i=1}^{10} \psi_{{\rm a},i} F_i, \quad \omega =  \sum_{i=1}^{10} \omega_{i} F_i,  \label{set_atm} \\
& \delta T_{\rm a} = \sum_{i=1}^{10} T_{\rm{a},i} F_i = 2 \frac{f_0}{R} \sum_{i=1}^{10} \theta_{{\rm a},i} F_i, \label{set_thermal} \\
& \quad \psi_{\rm o} = \sum_{i=1}^8 \psi_{{\rm o},i} \phi_i, \delta T_{\rm o} = \sum_{i=1}^8 T_{{\rm o},i} \phi_i, \label{set_oc} 
\end{align}
\end{subequations}
where $\theta_{{\rm a},i}= (\psi^1_{{\rm a},i}-\psi^3_{{\rm a},i})/2$ and  $\psi_{{\rm a},i}= (\psi^1_{{\rm a},i}+\psi^3_{{\rm a},i})/2$,
with $\psi^1_{\rm a}$ and $\psi^3_{\rm a}$ 
the streamfunctions in the upper and lower layer of the atmosphere. 

The projection of the dynamical equations~\eqref{eq:atmos}--\eqref{eq:stress}
on the modes kept in Eqs.~\eqref{eq:atmos_set} and \eqref{eq:ocean_set}
has already been discussed in detail in
\cite{Reinhold1982, Vannitsem2014}, and it will not be repeated here. 
We describe, however, in detail the procedure for projecting the temperature equations.

The temperature equation \eqref{eq:heat_oc} within the ocean can 
be written as
\begin{eqnarray}\label{eq:Tko}
\frac{\partial T_{{\rm o},k}}{\partial t} + \sum_i \sum_j \psi_{{\rm o},i} T_{{\rm o},j} d_{k,i,j} & = & - \frac{\lambda}{\gamma_{\rm o}} (T_{{\rm o},k} -
2 \frac{f_0}{R} \sum_i \theta_{{\rm a},i} e_{k,i}) - \frac{4}{\gamma_{\rm o}} \sigma_B T_{{\rm o},0}^3 T_{{\rm o},k} \nonumber \\
& + & \frac{8 \epsilon_{\rm a}}{\gamma_{\rm o}} \sigma_B T_{{\rm a},0}^3 \frac{f_0}{R} \sum_i \theta_{{\rm a},i} e_{k,i}  
+ \frac{C_{\rm o}}{\gamma_{\rm o}} e_{k,1},
\end{eqnarray}
where
\begin{subequations}\label{eq:temp_set_oc_de}
\begin{align}
& d_{k,i,j} = \frac{n}{2 \pi^2 L^2} \int_0^{\pi L} \int_0^{2\pi L/n} dy\, dx \, \phi_k \, J(\phi_i,\phi_j),  \label{coeffs_adv}  \\
& e_{k,i} = \frac{n}{2 \pi^2 L^2} \int_0^{\pi L} \int_0^{2\pi L/n} dy\, dx\, \phi_k \, F_i.  \label{coeffs_lin}
\end{align}
\end{subequations}

Non-dimensionalizing the baroclinic and barotropic streamfunction fields,
$\theta'_{{\rm a},i}= \theta_{{\rm a},i}/(L^2 f_0)$ and $\psi'_{{\rm a},i}= \psi_{{\rm a},i}/(L^2 f_0)$,
and the parameters of the truncated, low-order model, such that
$\lambda'=\lambda/(\gamma_{\rm o} f_0)$, $\sigma'_{B,{\rm a}} = 8 \epsilon_{\rm a} \sigma_B T_{{\rm a},0}^3 / (\gamma_{\rm o} f_0)$,
$\sigma'_{B,{\rm o}} = 4 \sigma_B T_{{\rm o},0}^3 / (\gamma_{\rm o} f_0)$, and $C'_{\rm o} = C_{\rm o}/ (\gamma_{\rm o} f_0)$, one 
gets
\begin{eqnarray}
\frac{\partial T_{{\rm o},k}}{\partial t} + \sum_i \sum_j \psi'_{{\rm o},i} T_{{\rm o},j} d'_{k,i,j} & = & - \lambda' (T_{{\rm o},k} -
2 \frac{f_0^2 L^2}{R} \sum_i \theta'_{{\rm a},i} e_{k,i}) - \sigma'_{B,{\rm o}} T_{{\rm o},k} \nonumber \\
& + & \sigma'_{B,{\rm a}} \frac{f_0^2 L^2}{R} \sum_i \theta'_{{\rm a},i} e_{k,i} + C'_{\rm o} e_{k,1},
\label{eq:temp_set_oc_de_1}
\end{eqnarray}
where $ d'_{k,i,j}=  n/(2 \pi^2) \int_0^{\pi} \int_0^{2\pi /n} dy'\,  dx'\,  \phi_k \, J'(\phi_i,\phi_j)$.
The temperature in the ocean can be further normalized by 
$(f_0^2 L^2)/R$.

For the atmosphere, the procedure is the same and one gets
\begin{eqnarray}
\frac{\partial \theta'_{{\rm a},k}}{\partial t} + \sum_i \sum_j \psi'_{{\rm a},i}  \theta'_{{\rm a},j} c'_{k,i,j} = 
\sigma' \omega'_k- \lambda'_{\rm a} ( \theta'_{{\rm a},k} -
\frac{1}{2} \frac{R}{L^2 f_0^2} \sum_i T_{{\rm o},i} s_{k,i}) \nonumber \\
+ S_{B,{\rm o}} \frac{R}{L^2 f_0^2} (\sum_i T_{{\rm o},i} s_{k,i}) - S_{B,{\rm a}} \theta'_{{\rm a},k}  
+ C'_{\rm a} \frac{R}{L^2 f_0^2} \delta^k_1,
\label{temp}
\end{eqnarray}
with $\sigma' = \sigma \delta p^2/(2 L^2 f_0^2)$, $\lambda'_{\rm a} = \lambda/(\gamma_{\rm a} f_0)$,
$S_{B,{\rm a}} = (8 \epsilon_{\rm a}/(\gamma_{\rm a} f_0) \sigma_B T_{{\rm a},0}^3$, $S_{B,{\rm o}} =  (4 \epsilon_{\rm a})/( 2 \gamma_{\rm a} f_0) \sigma_B T_{{\rm o},0}^3$, and $C'_{\rm a} = C_{\rm a}/(2 \gamma_{\rm a} f_0)$, 
while the projection coefficients are given by
\begin{subequations}\label{eq:temp_set_oc_cs}
\begin{align}
& c'_{k,i,j} = \frac{n}{2 \pi^2} \int_0^{\pi  } \int_0^{2\pi /n} dy'\,  dx'\,  F_k \, J'(F_i,F_j), \\
& s_{k,i} = \frac{n}{2 \pi^2 } \int_0^{\pi } \int_0^{2\pi /n} dy'\,  dx'\,  \phi_i \, F_k. 
\end{align}
\end{subequations}

The set of default
values of the nondimensional parameters is
given in Table~\ref{tab:nondim_param}. The
parameter $C_{\rm o}$ --- or, equivalently, $C'_{\rm a}$ and $C'_{\rm o}$ --- will be varied in the dynamical analysis of Section~\ref{sec:results}.
The default value for $\lambda$ appearing in the estimations of $\lambda'_{\rm o}$ and $\lambda'_{\rm a}$ will also be
varied at the end of the next section in order to clarify the impact of the heat flux coupling.


\newpage
\setcounter{figure}{0}
\renewcommand{\thefigure}{\arabic{figure}}
\setcounter{table}{0}
\renewcommand{\thetable}{\arabic{table}}

\begin{table}
\centering
\caption{Typical dimensional parameter values used in \cite{Barsugli1998}}
\label{tab:dim_param}     
\begin{tabular}{|l|}
\hline
Parameter values \\
\hline
$R_{\rm o}=200$ Wm$^{-2}$      \\
$\epsilon_{\rm a} = 0.76 $       \\
$\sigma_B = 5.6 \cdot 10^{-8}$  Wm$^{-2}$K$^{-4}$      \\
$\gamma_{\rm o} = 2 \cdot 10^{8}$  Jm$^{-2}$K$^{-1}$      \\
$\lambda = 20$ Wm$^{-2}$K$^{-1}$  \\
$\gamma_{\rm a} = 10^{7}$  Jm$^{-2}$K$^{-1}$  \\
$T_{{\rm a},0} = 270$ K  \\
$T_{{\rm o},0} = 285$ K  \\
\hline
\end{tabular}
\end{table}

\begin{table}
\centering
\caption{Default 
values of the nondimensional parameters in the temperature equations}
\label{tab:nondim_param}     
\begin{tabular}{|l|}
\hline
Parameter values \\
\hline
$\sigma'=0.1; \quad  \alpha= 1/4$ or $1/3$      \\
$S_{B,{\rm a}}= 6.49 10^{-3}; \quad S_{B,{\rm o}}= 3.82 10^{-3}$  \\
$\sigma'_{B,{\rm a}}=3.25 10^{-4};  \quad  \sigma'_{B,{\rm o}}=2.51 10^{-4}$  \\
$\lambda'_{\rm o}=9.69 10^{-4};  \quad  \lambda'_{\rm a}=1.94 10^{-2}$    \\
\hline
\end{tabular}
\end{table}

\begin{figure}
\centering
{\includegraphics[width=0.95\textwidth]{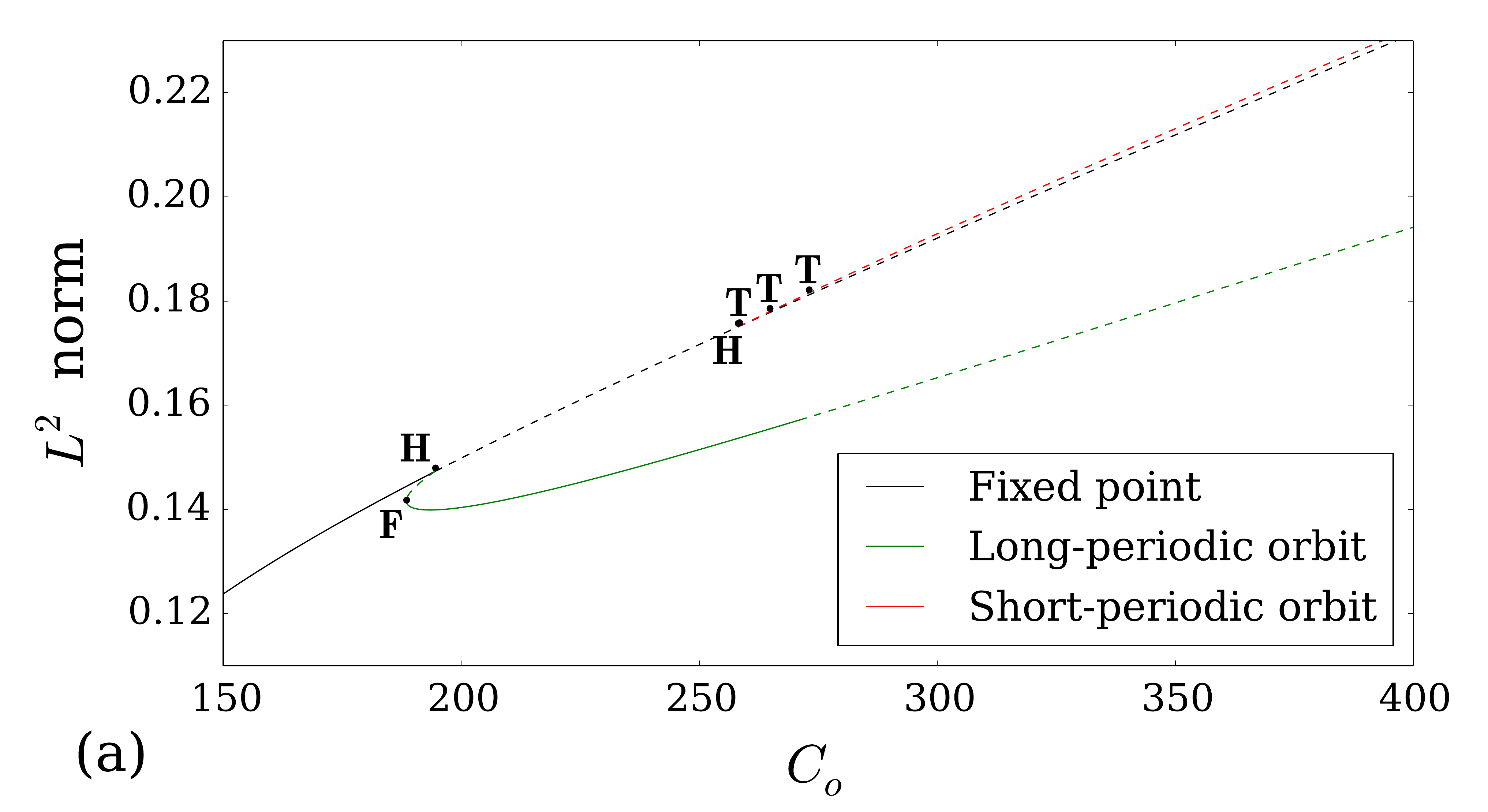}}
{\includegraphics[width=0.95\textwidth]{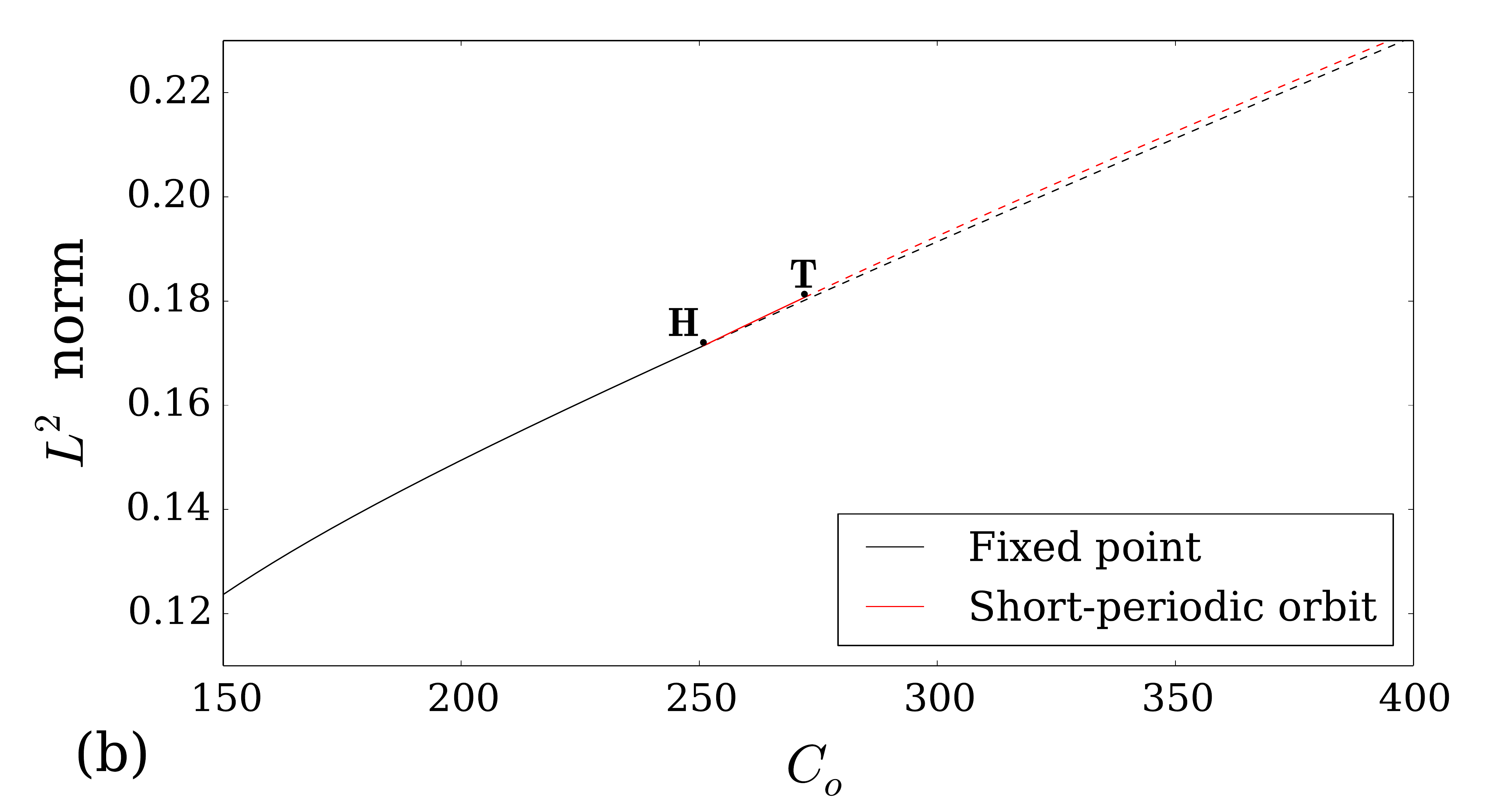}}
\caption{Successive bifurcations of the coupled model. 
(a) Bifurcation diagram obtained by varying the meridional change $C_{\rm o}$ in the radiative input from the Sun for a fixed value of 
the mechanical atmosphere-ocean coupling, $d=1 \times 10^{-8}$ s$^{-1}$. The 
$L^2$-norm of the solutions is plotted on the ordinate.
Two branches of periodic orbits emanate from the Hopf bifurcations (denoted by {\bf H}) 
that lie on the principal branch (black curve): a short-periodic one (red curve) and a long-periodic
one (green curve). The stability of the branches is indicated by solid (stable) or dashed (unstable) lines.
On the periodic-orbit branches, a fold ({\bf F} symbol) and torus ({\bf T} symbols) bifurcations are also 
shown.  (b) Bifurcation diagram obtained by varying $C_{\rm o}$ 
in the absence of wind friction, i.e. for 
zero mechanical coupling, $d = 0$. 
}
\label{fig:Hopf}
\end{figure}

\begin{figure}
\centering
{\includegraphics[width=0.95\textwidth]{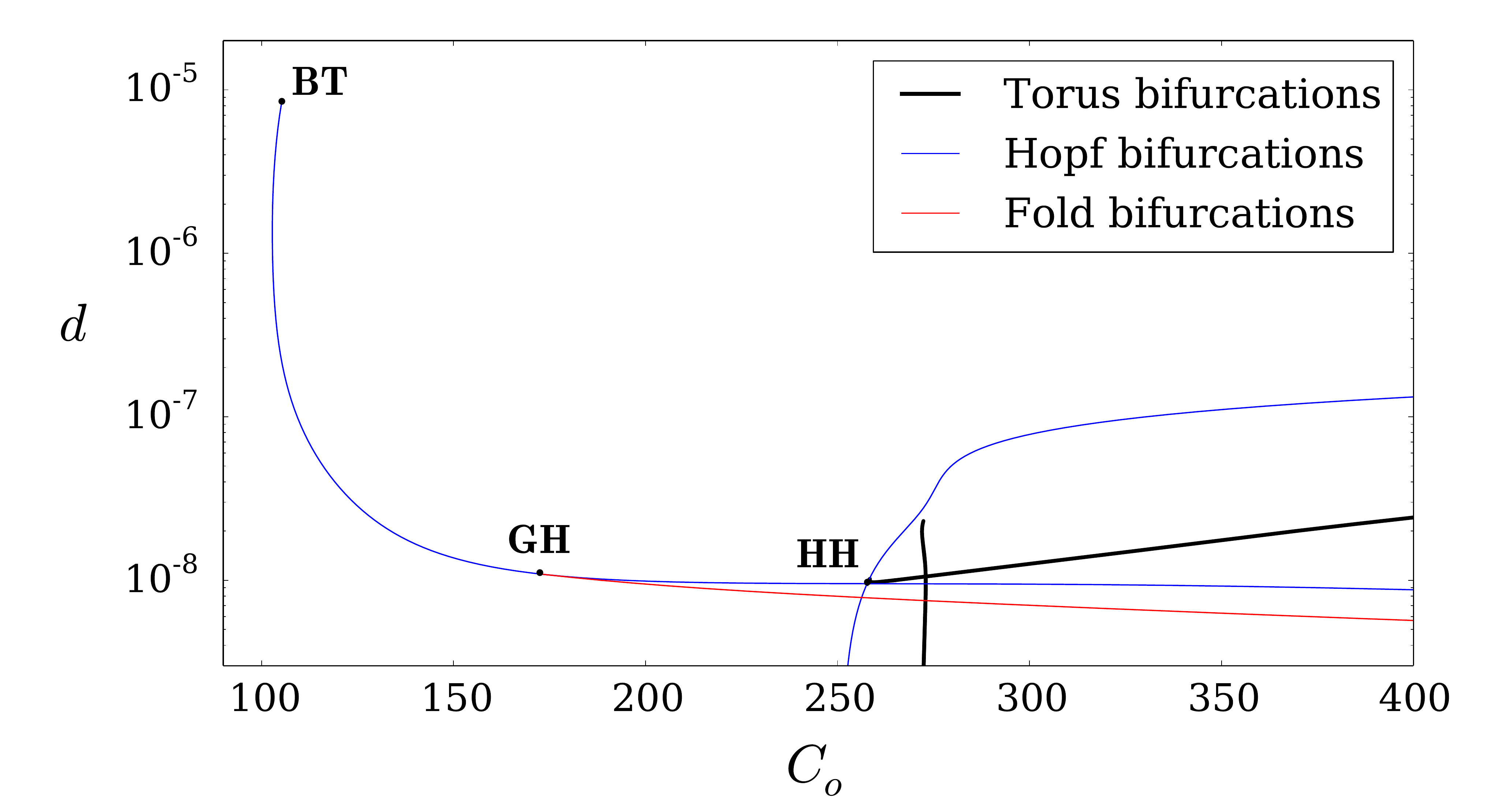}}
\caption{
A semi-logarithmic regime diagram in the $(C_{\rm o}, \log_{10}d)$-plane:  
curves indicate the locus of codimension-1 bifurcations;  
their intersections are codimension-2 bifurcations ---  
Hopf--Hopf bifurcation ({\bf HH} symbol) at the intersection between the loci of the two Hopf bifurcations; the generalized Hopf bifurcation ({\bf GH} symbol), 
from which the locus of a fold emanates and the Bogdanov-Takens bifurcation ({\bf BT} symbol) from which the locus of the Hopf bifurcation that 
generates the long-periodic branch originates.} 
\label{fig:regime}
\end{figure}

\begin{figure}
\centering
{\includegraphics[width=0.95\textwidth]{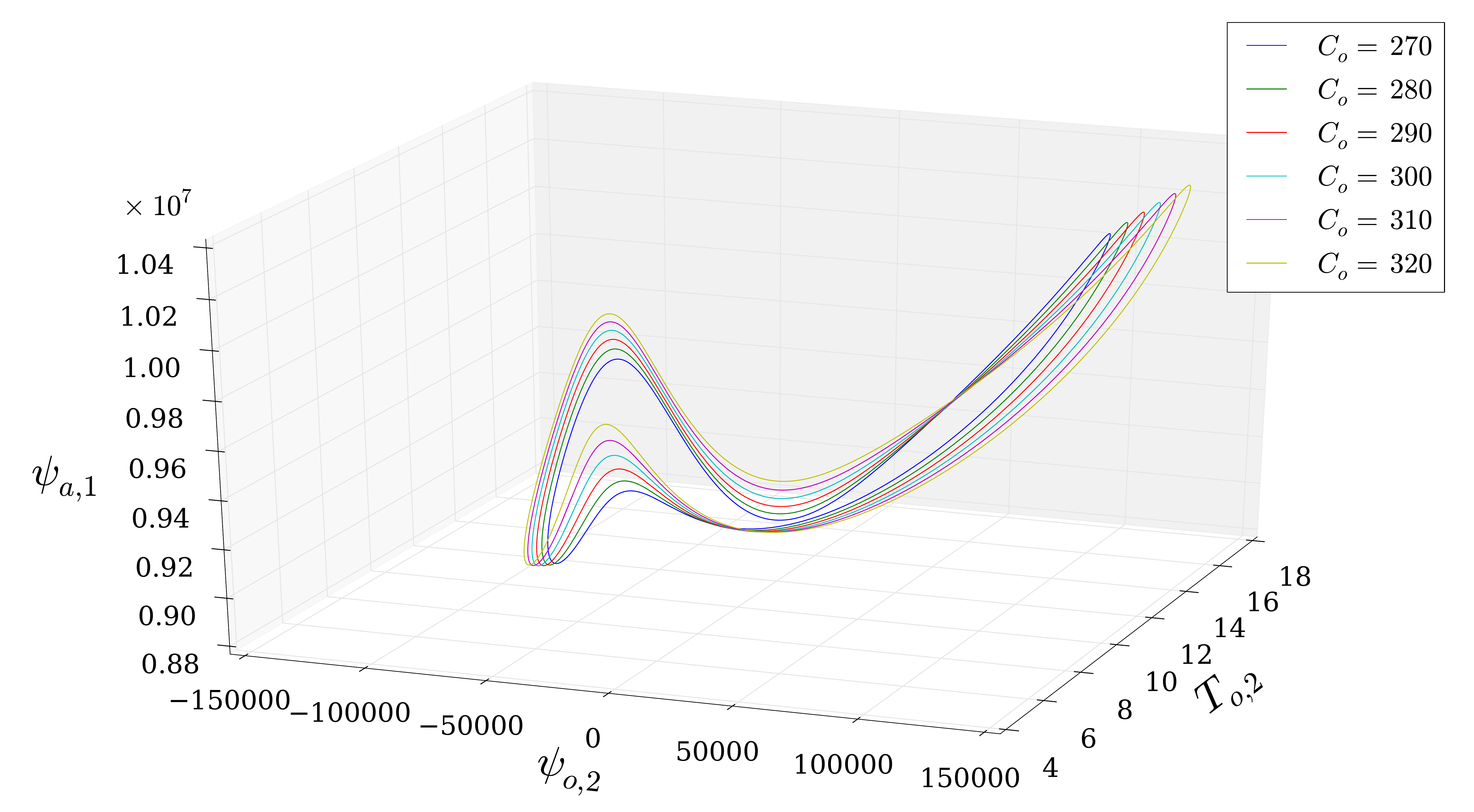}}
\caption{Long-periodic orbits of the coupled model, for
        several values of $C_{\rm o}$ (see legend), with $d=1 \times 10^{-8}$ s$^{-1}$. Plotted is a three-dimensional
        (3-D) projection of these orbits onto the modes $(\psi_{{\rm a},1}, \psi_{{\rm o},2}, T_{{\rm o},2})$.}
\label{fig:LCs}
\end{figure}

\begin{figure}
\centering
{\includegraphics[width=0.48\textwidth]{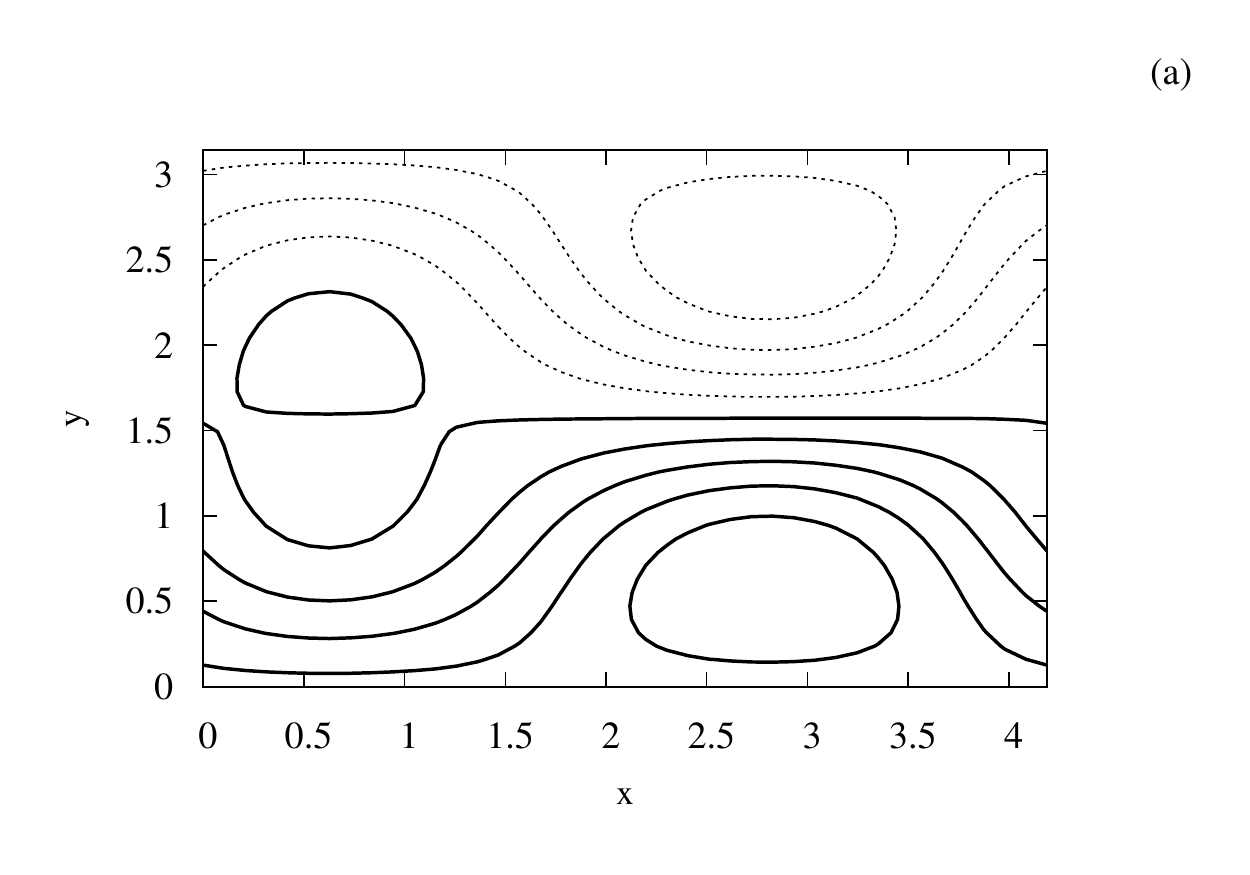}}
{\includegraphics[width=0.48\textwidth]{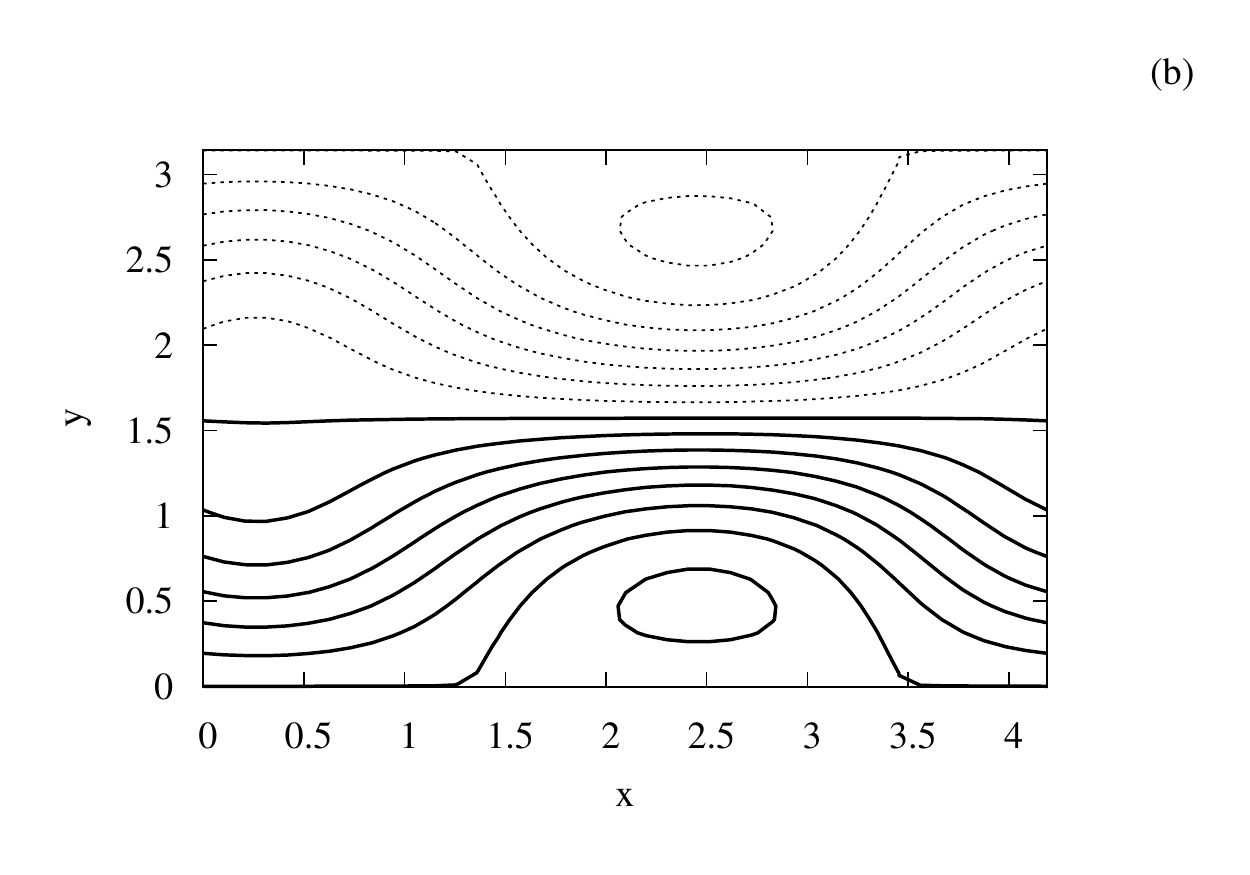}}
{\includegraphics[width=0.48\textwidth]{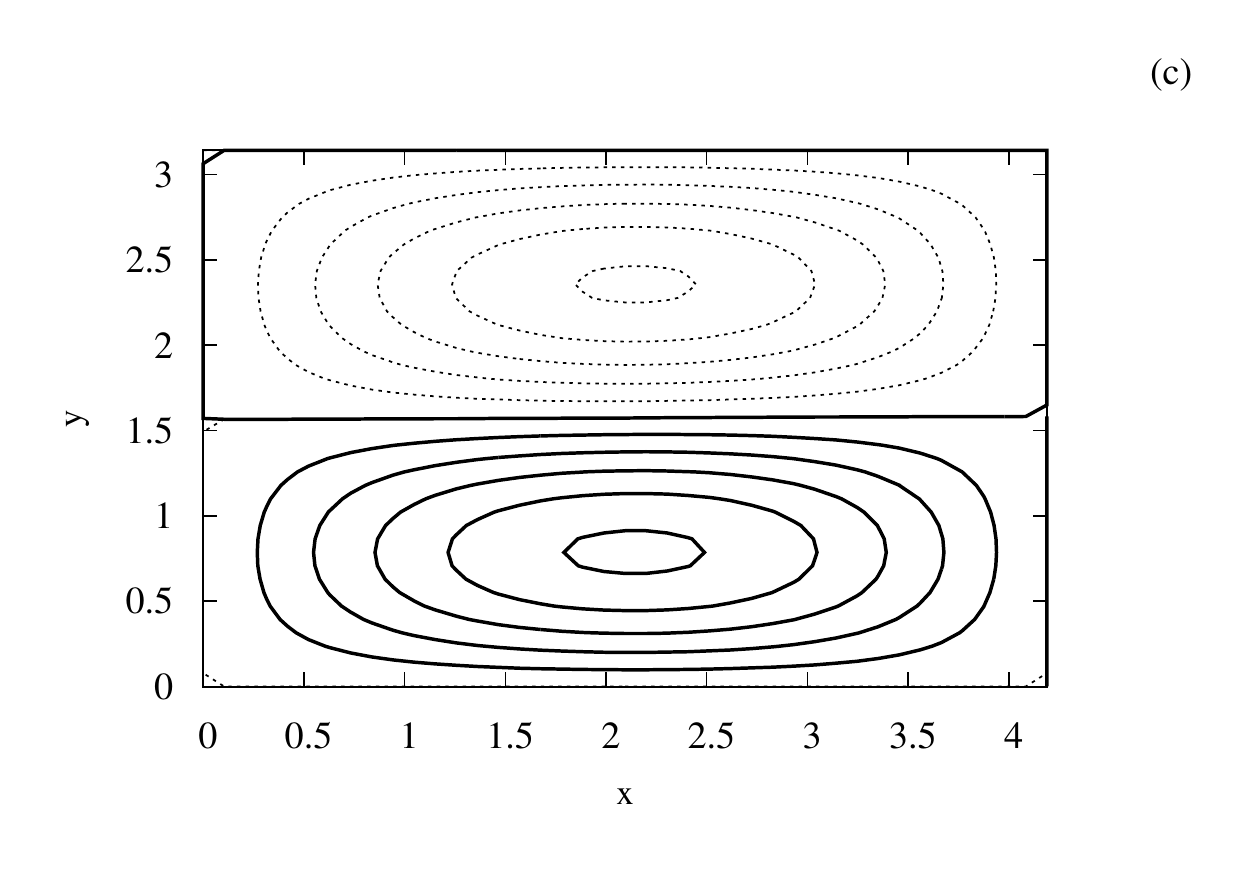}}
{\includegraphics[width=0.48\textwidth]{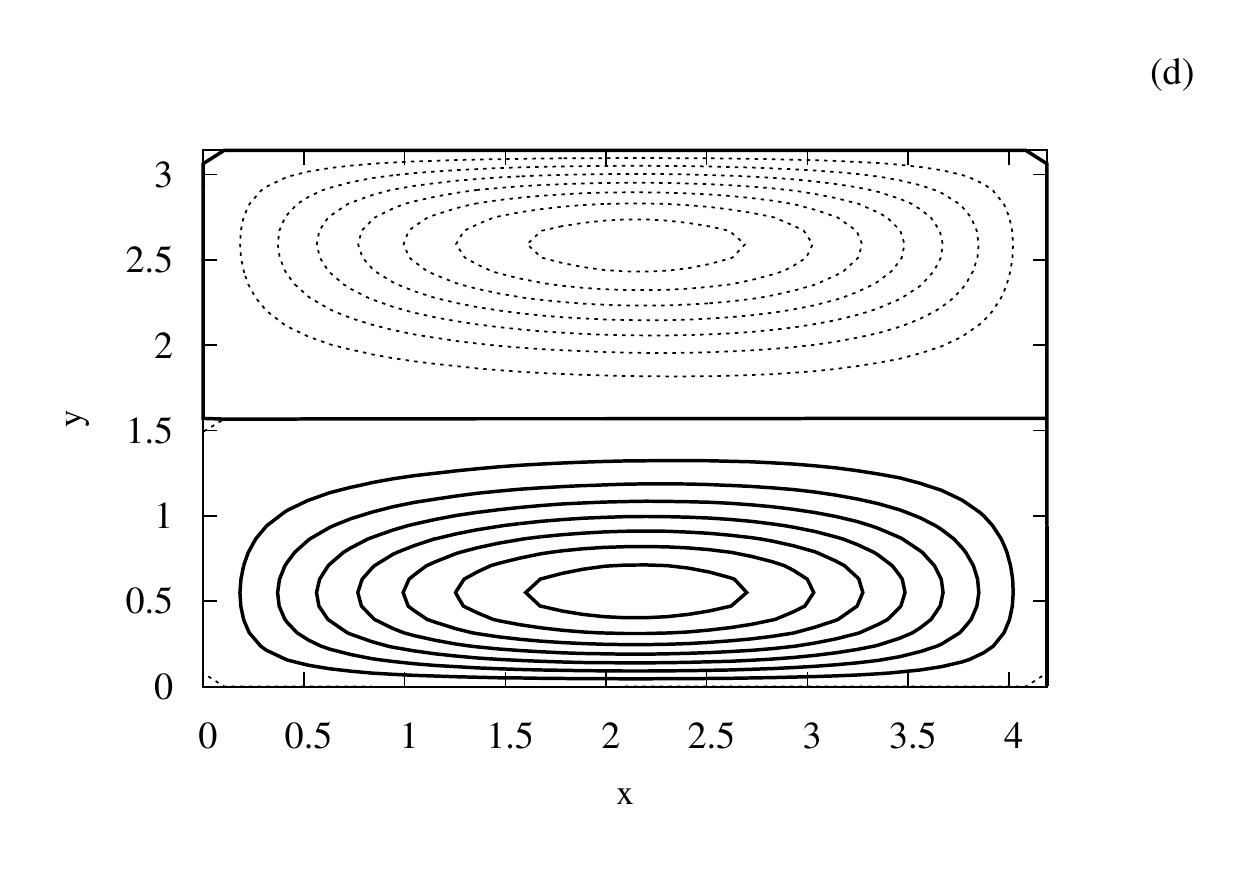}}
\caption{Climatology of the coupled model, at fixed parameter values. 
   Temporal averages over about $1.4 \times 10^{6}$ days days of (a) the atmospheric barotropic streamfunction; 
   (b) the atmospheric temperature anomaly; (c) the ocean streamfunction; and (d) the ocean temperature anomaly, 
   for $C_{\rm o}=300$ Wm$^{-2}$ and  $d=10^{-8}$ s$^{-1}$. Space coordinates are in nondimensional units. 
   The contour lines are solid for positive values and dashed for the negative ones; contour intervals (CI) 
   are uniform in all panels. 
   The CI and ranges (R) equal: (a) CI $= 0.5 \times 10^7$, R $= (-2.5, 2.5) \times 10^7$ m$^2$s$^{-1}$;  
   (b) CI $= 2$, R = (-15, 15) K; (c) CI $= 0.5 \times 10^4$, R $= (-3, 3) \times 10^4$ m$^2$s$^{-1}$; and 
   (d) CI $= 5$, R $= (-40, 40)$ K. 
   }
\label{fig:mean}
\end{figure}

\begin{figure}
\centering
{\includegraphics[width=0.48\textwidth]{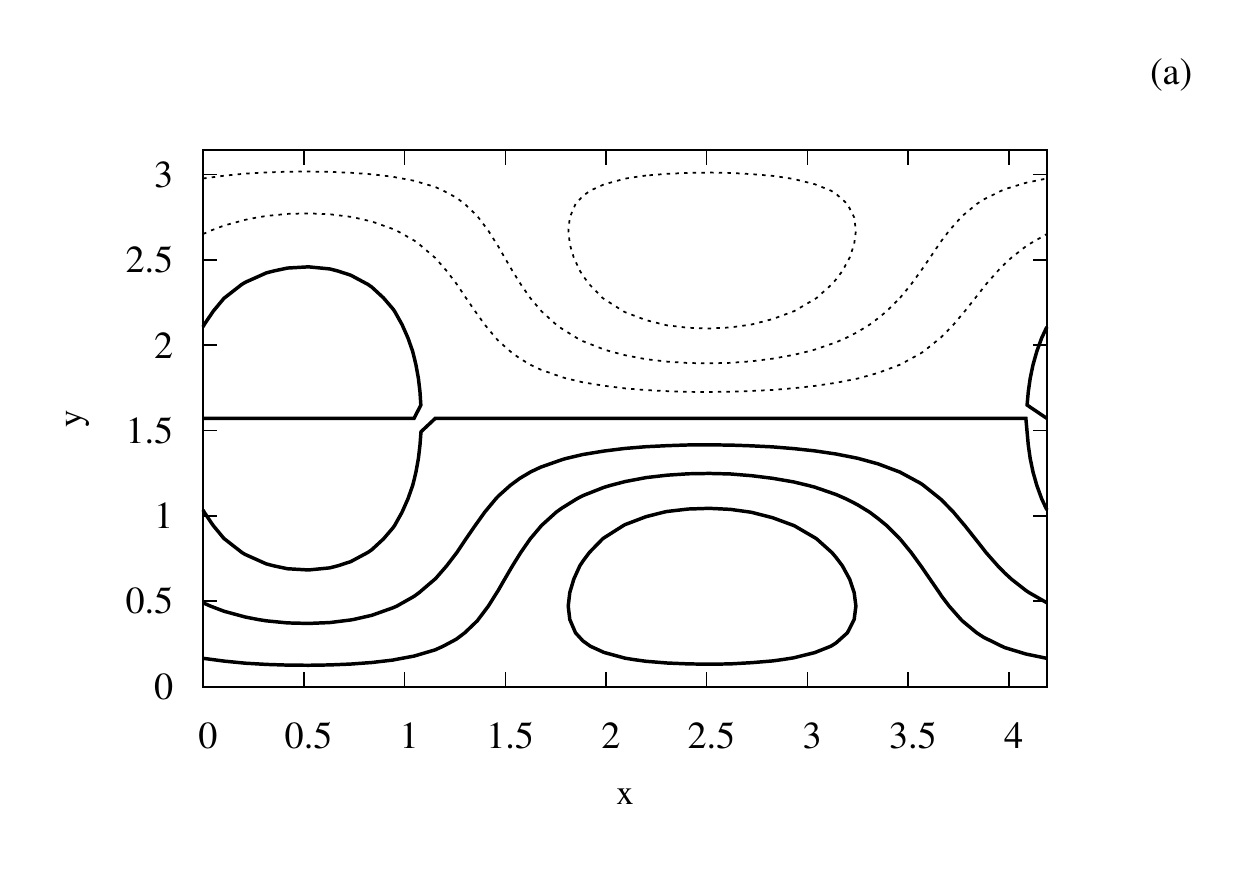}}
{\includegraphics[width=0.48\textwidth]{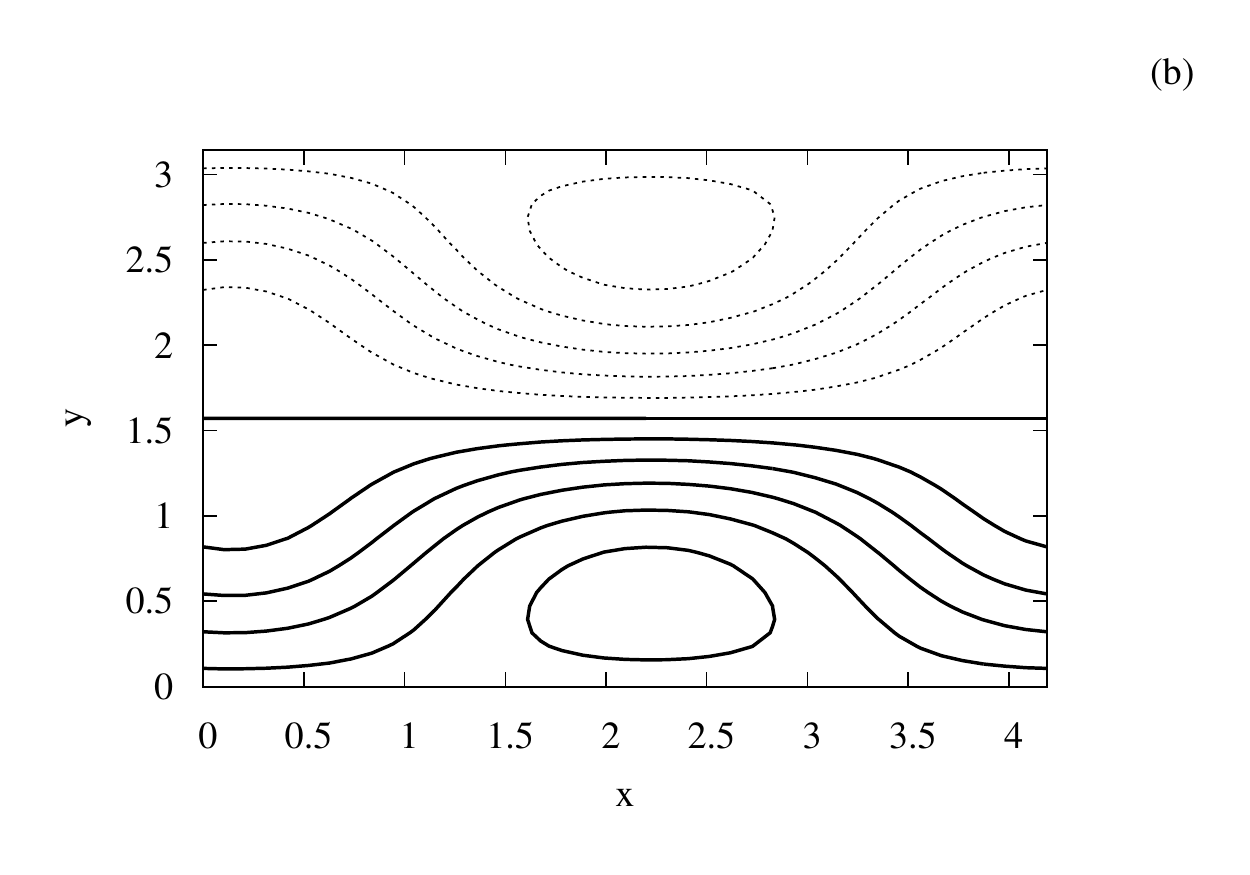}}
{\includegraphics[width=0.48\textwidth]{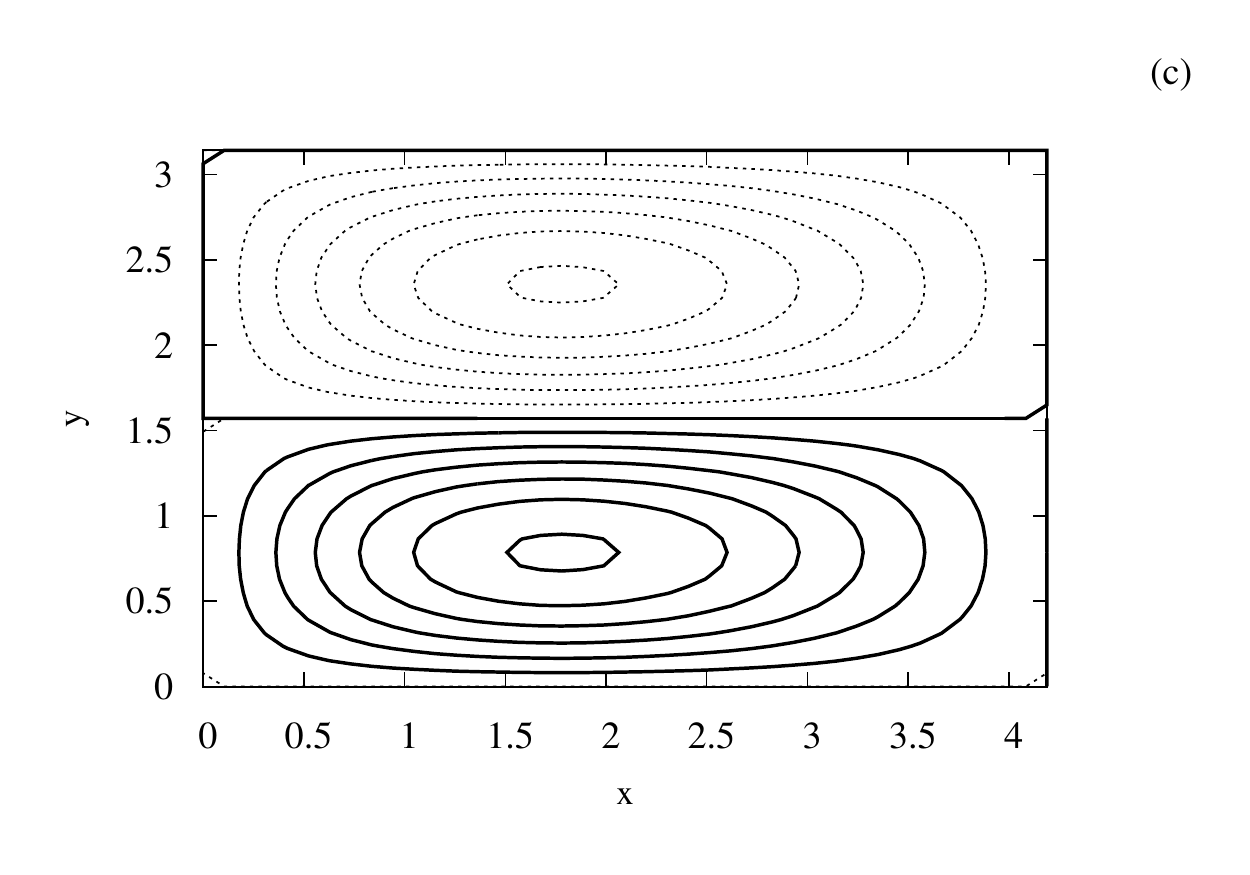}}
{\includegraphics[width=0.48\textwidth]{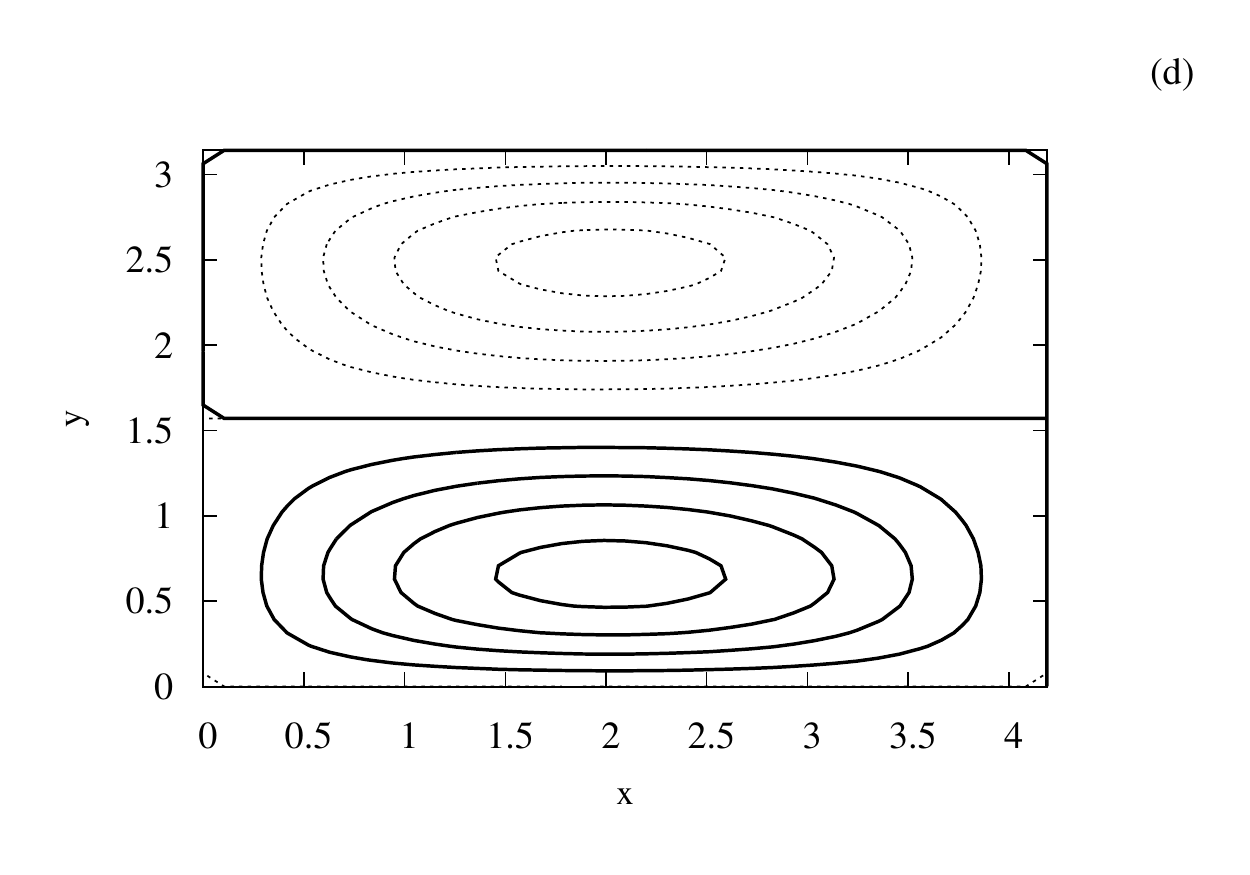}}
\caption{Same as Fig.~\ref{fig:mean}, but for
	$d=3.0 \times 10^{-8} s^{-1}$.
	The CI and R are fixed to (a) CI=$0.5 \times 10^{7}$, R$= (-2, 2) \times 10^7$ m$^2$s$^{-1}$, (b) CI=$2$ R=$ (-12, 12)$ K, (c) CI=$1 \times 10^{4}$, R$= (-7, 7) \times 10^4$ m$^2$s$^{-1}$, 
and (d) CI=$5$, R$= (-25, 25)$ K. 
	Note that the ranges of values are drastically different to the ones in Fig. \ref{fig:mean}.} 
\label{fig:couple_2} 
\end{figure}

\begin{figure}
\centering
{\includegraphics[width=0.9\textwidth]{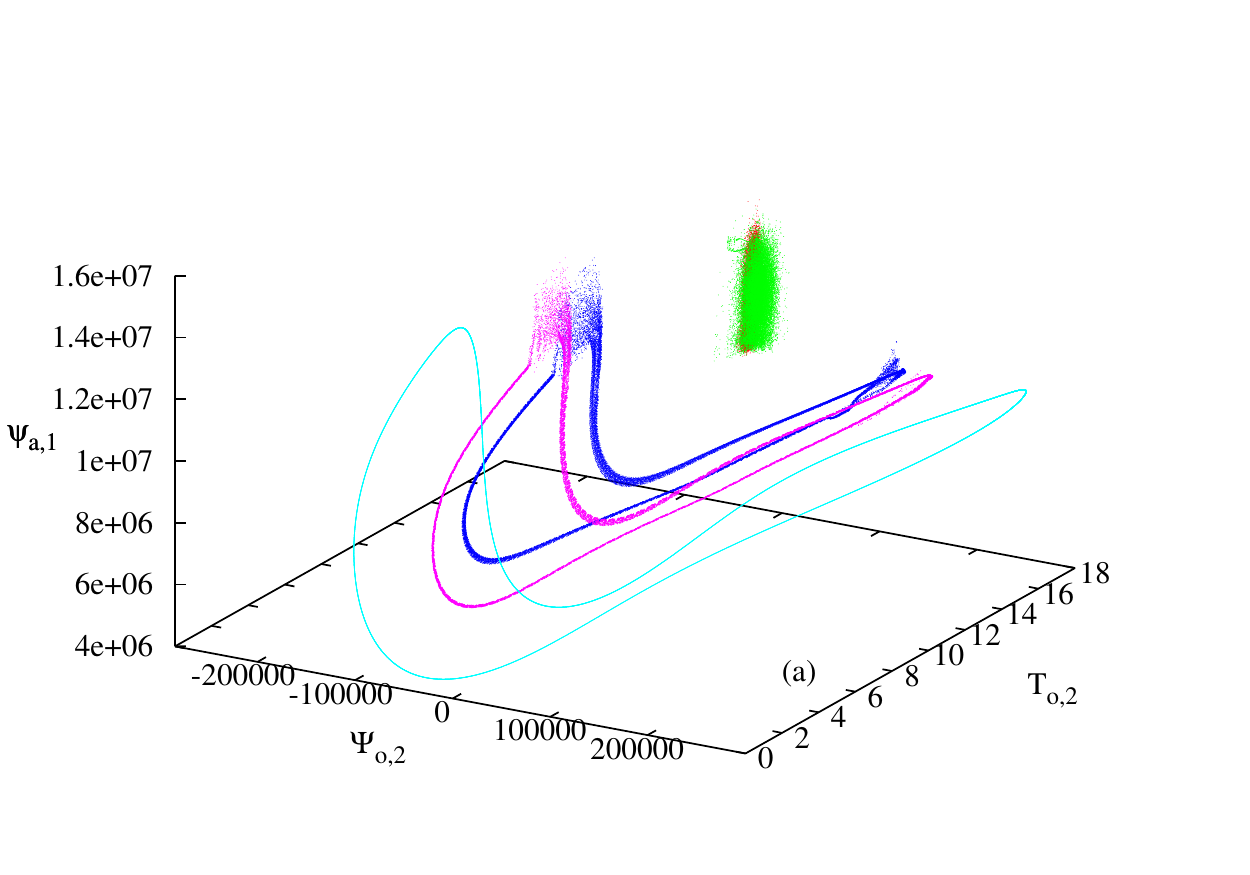}}
{\includegraphics[width=0.9\textwidth]{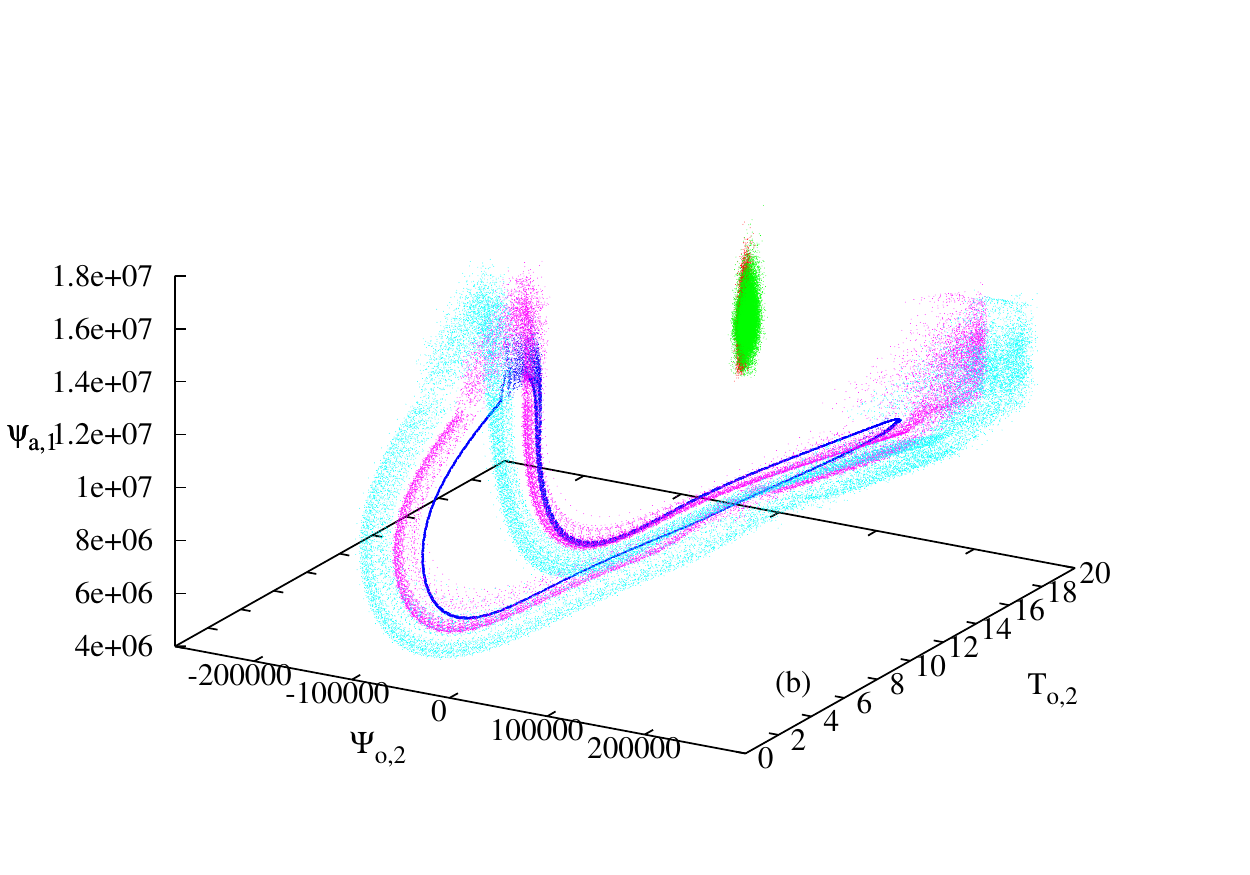}}
\caption{ 
   Similar to Fig.~\ref{fig:LCs}, but for several other parameter values. 
    (a) $C_{\rm o} = 300$ Wm$^{-2}$ and 
    several values of the friction parameter $d$: $5 \times 10^{-9}$ s$^{-1}$ (red),  
    $10^{-8}$ s$^{-1}$ (green), $2 \times 10^{-8}$ s$^{-1}$ (dark blue), 
    $3 \times10^{-8}$ s$^{-1}$ (magenta), and $8 \times 10^{-8}$ s$^{-1}$ (light blue); and 
    (b) for $C_{\rm o} = 350$ Wm$^{-2}$ with $d$: $5 \times10^{-9}$ s$^{-1}$ (red), $10^{-8}$ s$^{-1}$ (green), 
    $3 \times 10^{-8}$ s$^{-1}$ (dark blue), $5 \times 10^{-8}$ s$^{-1}$ (magenta), and 
    $8 \times 10^{-8}$ s$^{-1}$ (light blue). 
    }
\label{fig:slow_mfd}
\end{figure}

\begin{figure}
\centering
{\includegraphics[width=0.95\textwidth]{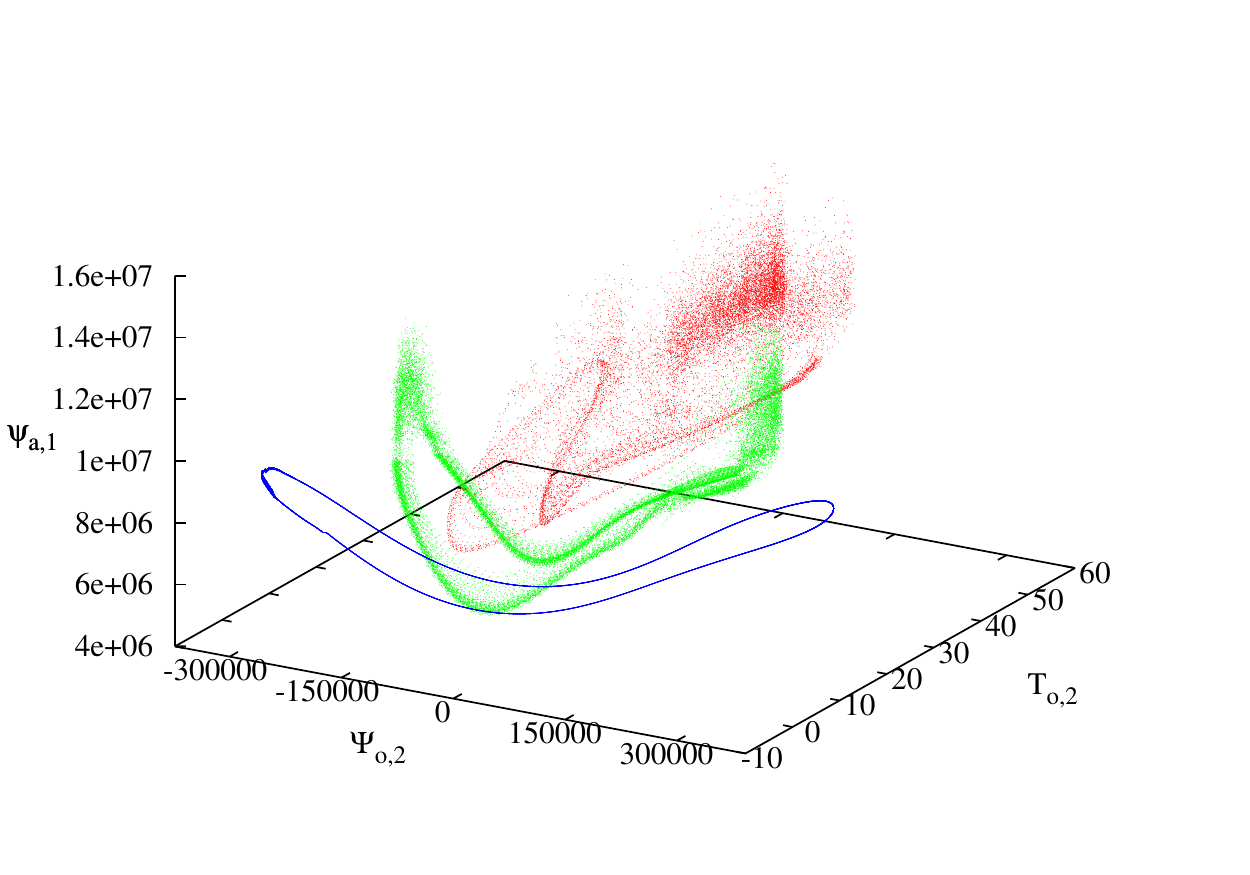}}
\caption{
	Same as Fig.~\ref{fig:slow_mfd} but for $C_{\rm o} = 350$ Wm$^{-2}$ and 
	$d=6 \times 10^{-8}$ s$^{-1}$, with $\lambda$: 0 (red), 20 (green), and 100 (blue), all in Wm$^{-2}$K$^{-1}$. 
	The dimensions of the associated attractors are discussed in Section~\ref{ssec:Lyap} below.
}
\label{fig:slow_lambda}
\end{figure}

\begin{figure}
\centering
{\includegraphics[width=0.9\textwidth]{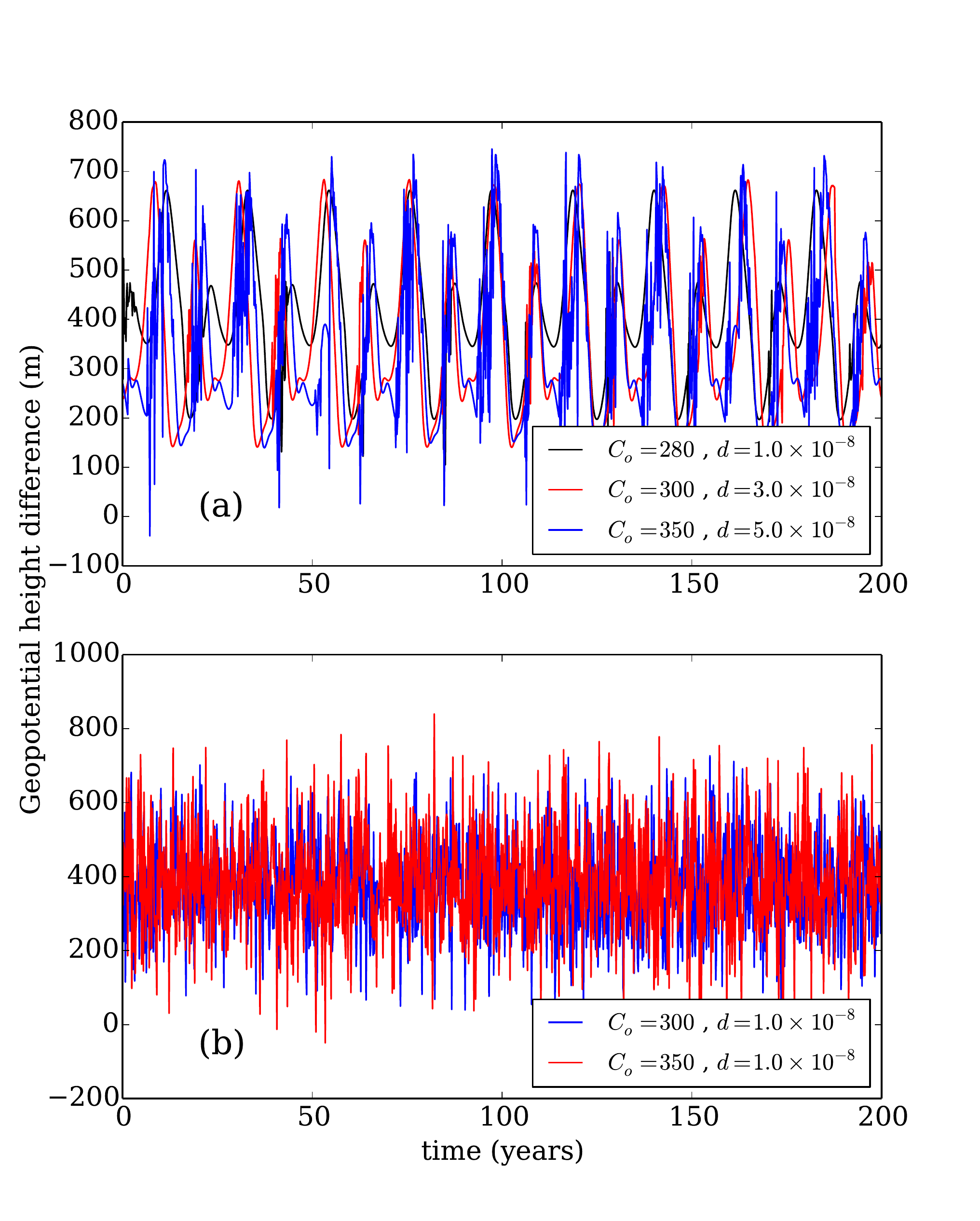}}
\caption{Low-frequency variability (LFV) of the coupled model. Time series of geopotential height difference 
    between locations ($\pi/n, \pi/4)$) and ($\pi/n, 3 \pi/4)$) of the model's nondimensional domain, for different values of
    meridional temperature gradient $C_{\rm o}$ and coupling coefficient $d$: 
    (a) chaotic but smooth trajectories living on a hypothetical slow attractor; 
    and (b) strongly fluctuating trajectories that are not lying close to such a slow attractor.
    }
\label{fig:LFV}
\end{figure}

\begin{figure}
\centering
\includegraphics[width=0.9\textwidth]{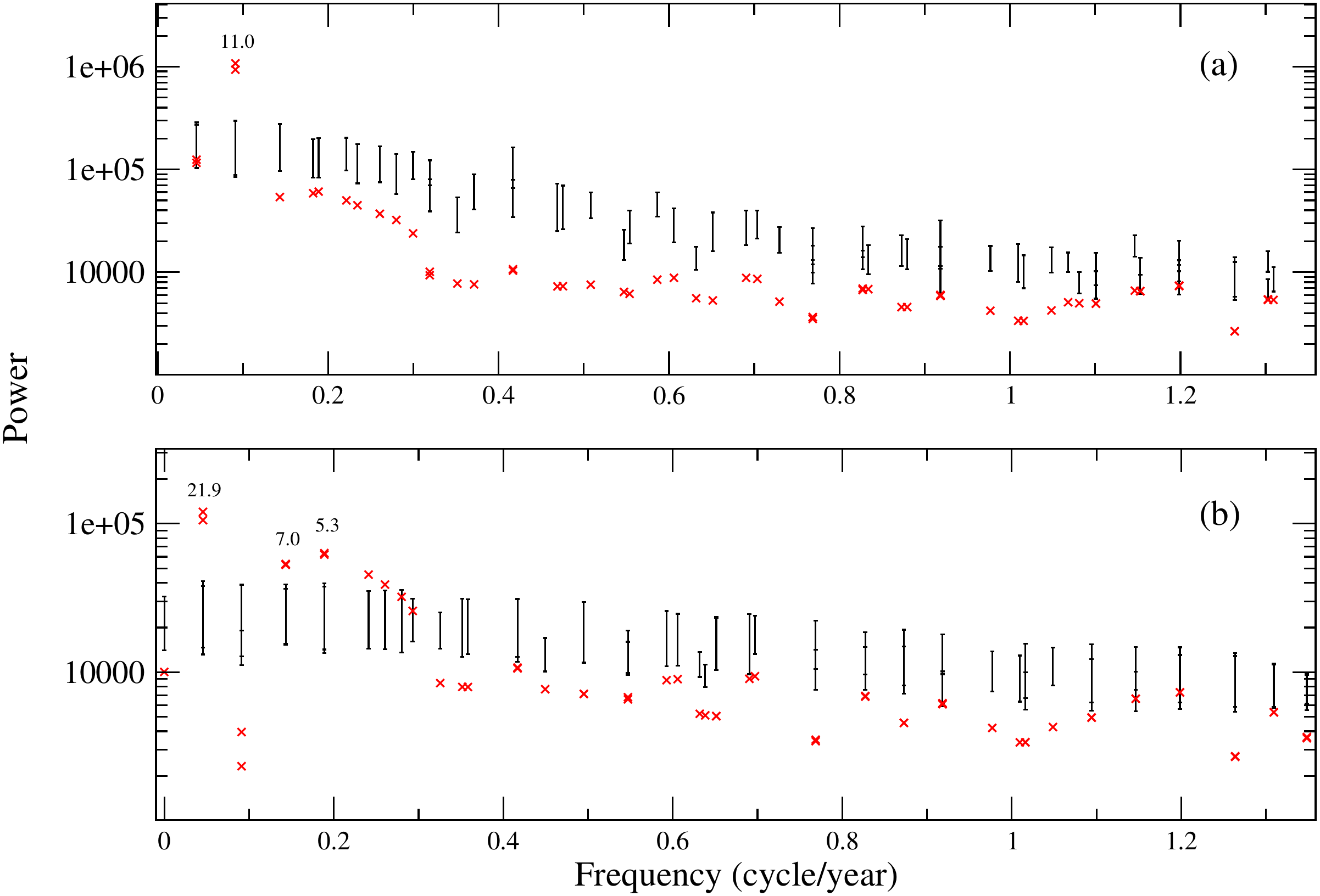}
\caption{Spectral analysis of
    a model simulation for $C_{\rm o}=350$ Wm$^{-2}$ and $d=5 \times 10^{-8}$s$^{-1}$; this simulation extends 
   the blue curve in Fig.~\ref{fig:LFV}a out to 500 years. The analysis uses Monte Carlo 
   SSA \cite{Allen1996, Ghil_al2002}, with a window width of $M = 30.7$ years. The estimated variance of each 
   mode in the data set is shown as a red cross, while lower and upper ticks on the errors bars 
   indicate the two-sided 95\% confidence interval based on an ensemble of 1000 red-noise surrogates; 
   see also \cite{FGR'11}. (a) Analysis for the full signal, showing a statistically significant pair of eigenvalues
   associated with the decadal time scale. A bidecadal time scale is also detected but it is not significant at the 
   95\% confidence level. (b) Same analysis, but after subtracting the decadal cycle. The bidecadal time scale 
   is now significant, according to this analysis, cf.~\cite{Allen1996, Ghil_al2002}.}
\label{fig:SSA}
\end{figure}

\begin{figure}
\centering
{\includegraphics[width=0.9\textwidth]{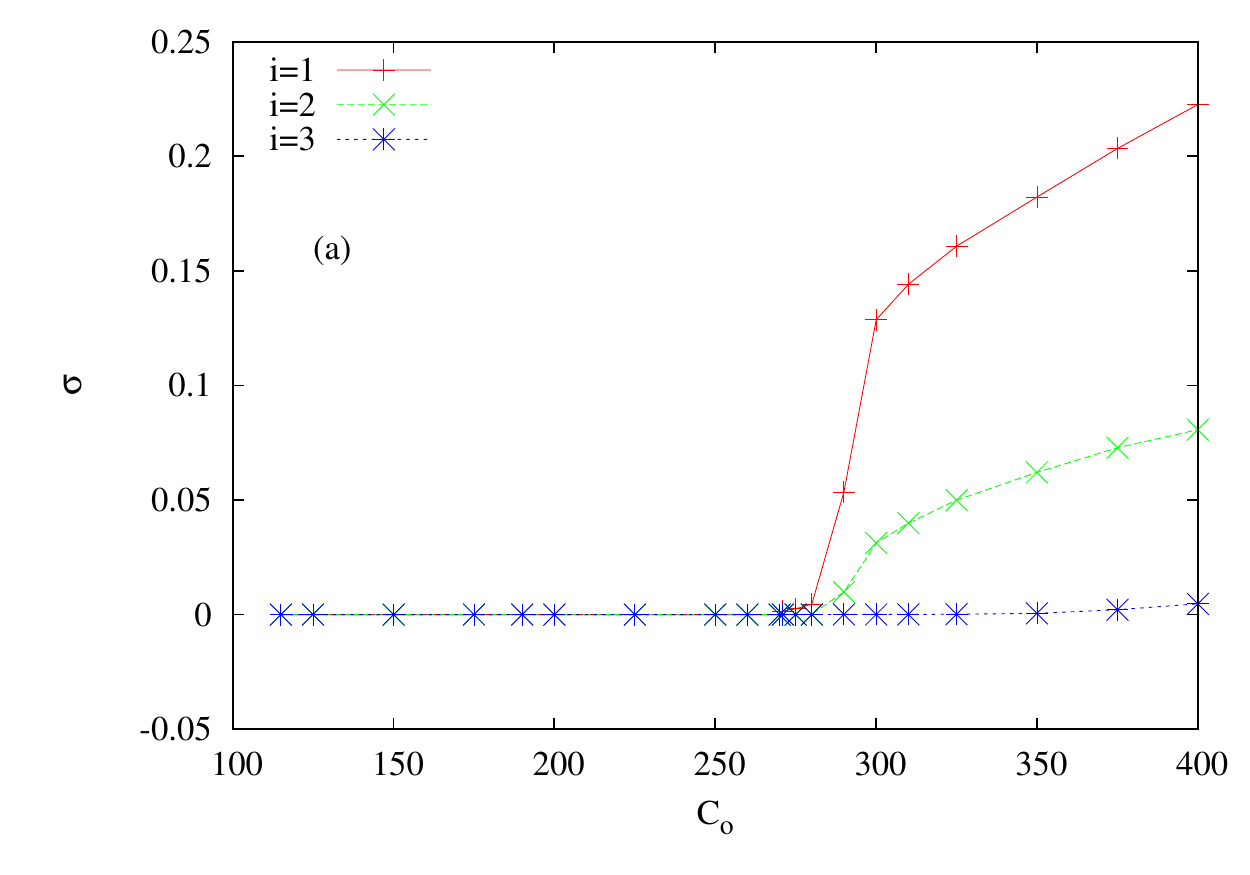}}
{\includegraphics[width=0.9\textwidth]{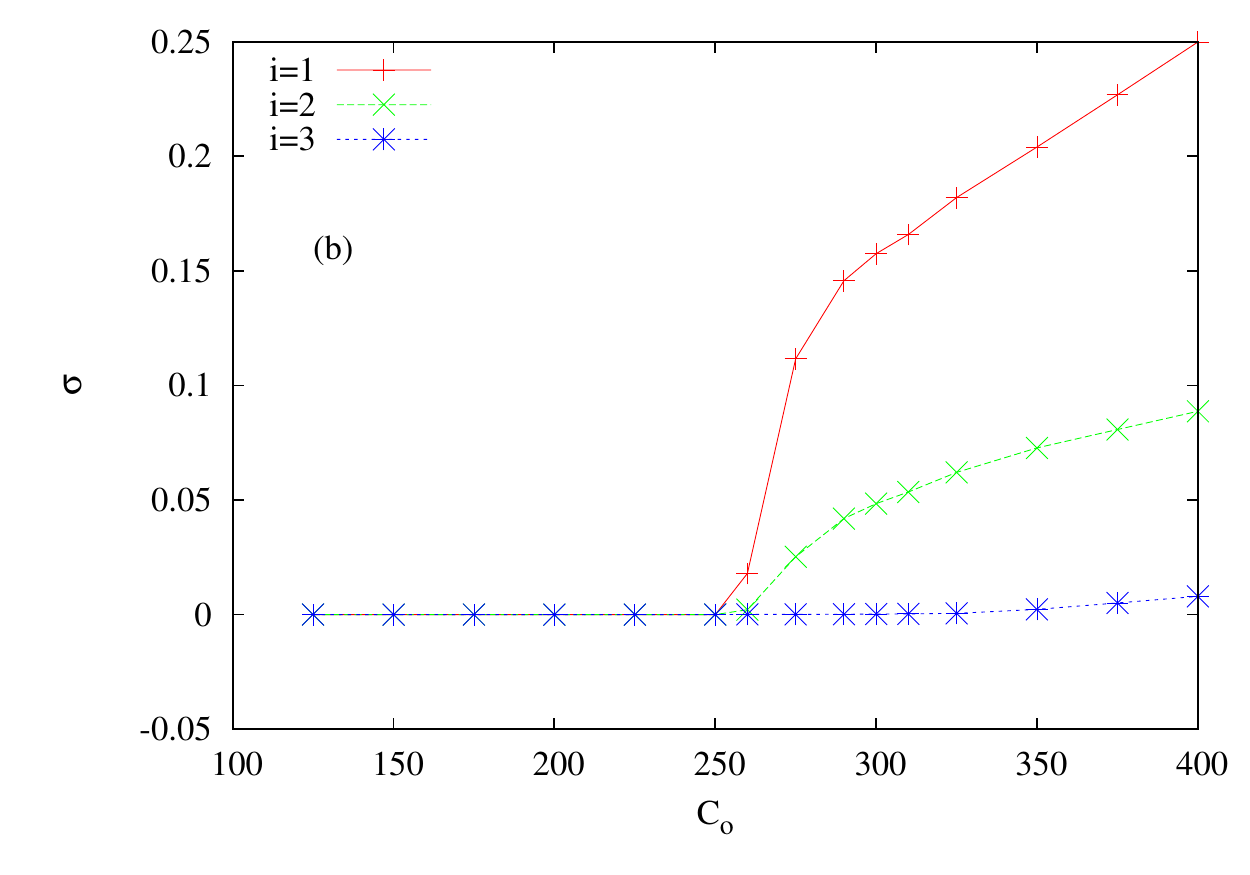}}
\caption{ Leading Lyapunov exponents 
as a function of $C_{\rm o}$. Values are given in day$^{-1}$ for the first three exponents, $i = 1,2, 3$ in the legend (red, green and blue curves), and $d=10^{-8}$ s$^{-1}$ in both panels: 
(a) $C_{\rm a}=C_{\rm o}/4$; and (b) $C_{\rm a}=C_{\rm o}/3$. The mean values, here and in the following Figs.~\ref{fig:Lyap_spectrum}--\ref{fig:LyaP_heat},  were obtained after $1.4 \times 10^{6}$ days.}
\label{fig:lead_Lyap}
\end{figure}

\begin{figure}
\centering
{\includegraphics[width=100mm]{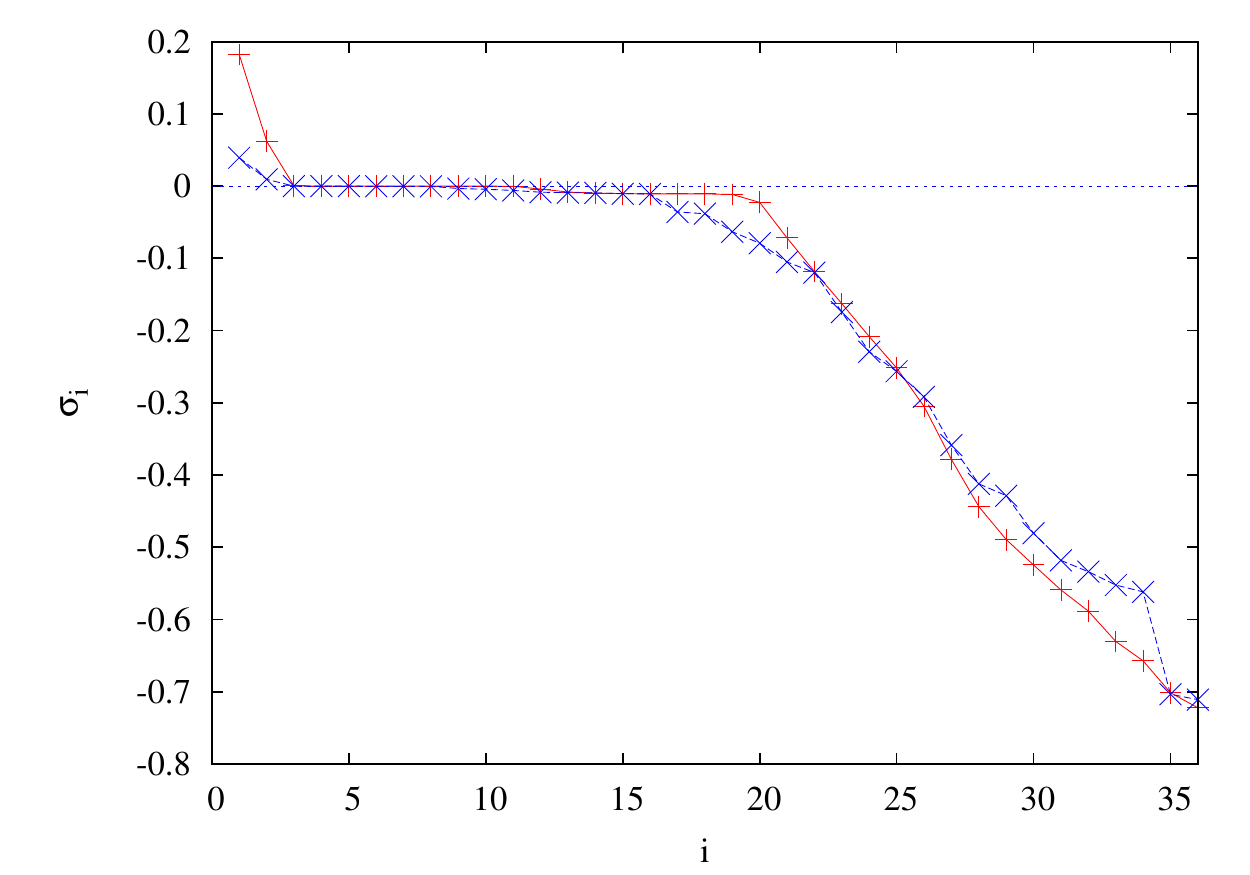}}
\caption{Full Lyapunov spectra  for $C_{\rm o}=350$ Wm$^{-2}$ and $C_{\rm a}=C_{\rm o}/4$, using an integration of $1.4 \times 10^{6}$ days;
red curve for $d=10^{-8}$ s$^{-1}$ 
and blue curve for
$d=5 \times 10^{-8}$ s$^{-1}$. 
}
\label{fig:Lyap_spectrum}
\end{figure}

\begin{figure}
\centering
{\includegraphics[width=100mm]{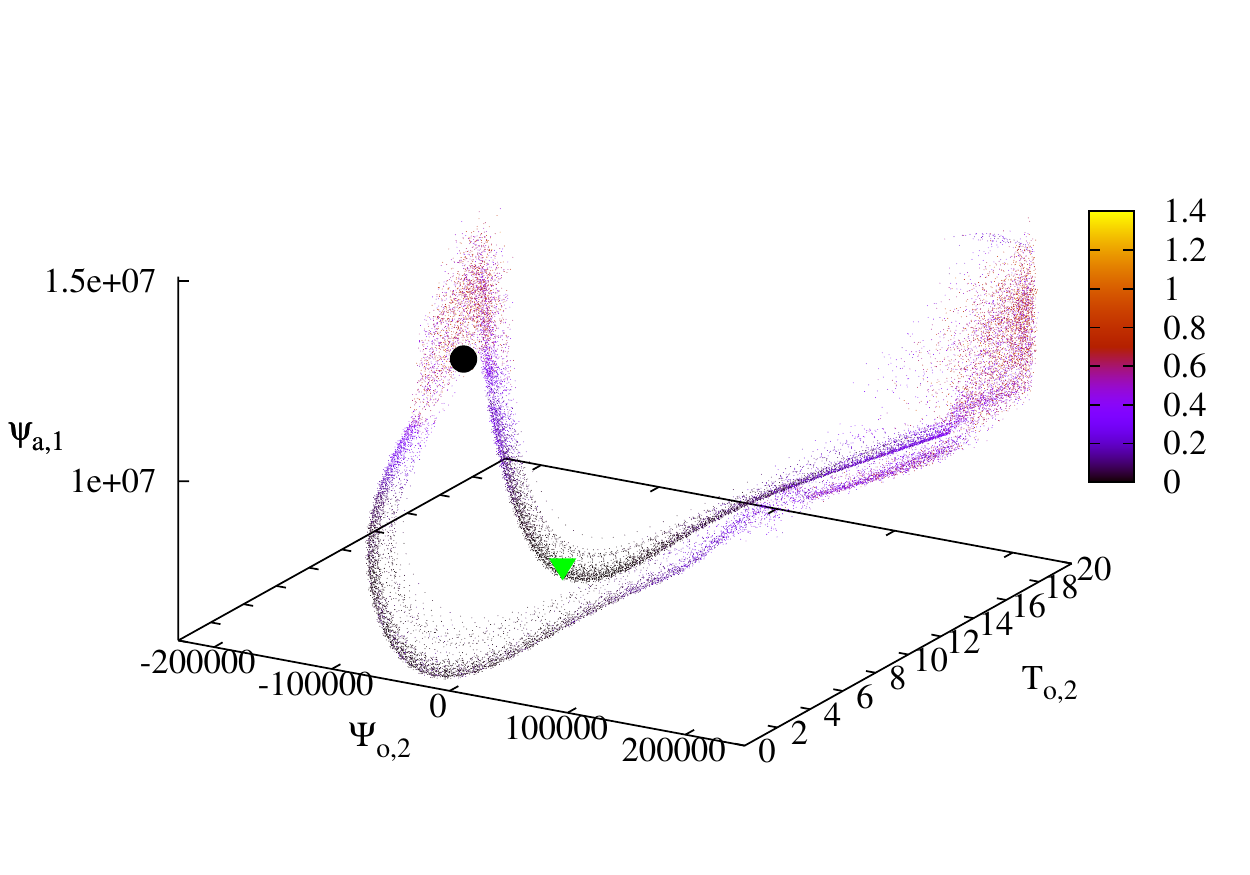}}
\caption{ 3-D plot of the local Kolmogorov-Sinai entropy $h_{\rm KS}(\x)$ for $C_{\rm o} = 350$~Wm$^{-2},$ $C_{\rm a} = C_{\rm o}/4$, 
   $d=5 \times 10^{-8}$~s$^{-1}$ and $\lambda = 20$~Wm$^{-2}$K$^{-1}.$ These parameter 
   values correspond to the blue curve    in Fig.~\ref{fig:Lyap_spectrum}, and the color bar shows that, in this case, $0 \leq h_{\rm KS}(\x) \leq 1.4$, 
   with very small values concentrated in the vicinity of slow and periodic --- or at least nearly periodic --- solutions.
    The green triangle indicates the approximate location of the initial states for the ensemble whose error 
    evolution is given by the green curve in Fig.~\ref{fig:predict} below, the 
    filled black circle indicates the approximate start for the runs associated with the black curve in that figure.}
\label{fig:KS}
\end{figure}

\begin{figure}
\centering
{\includegraphics[width=100mm]{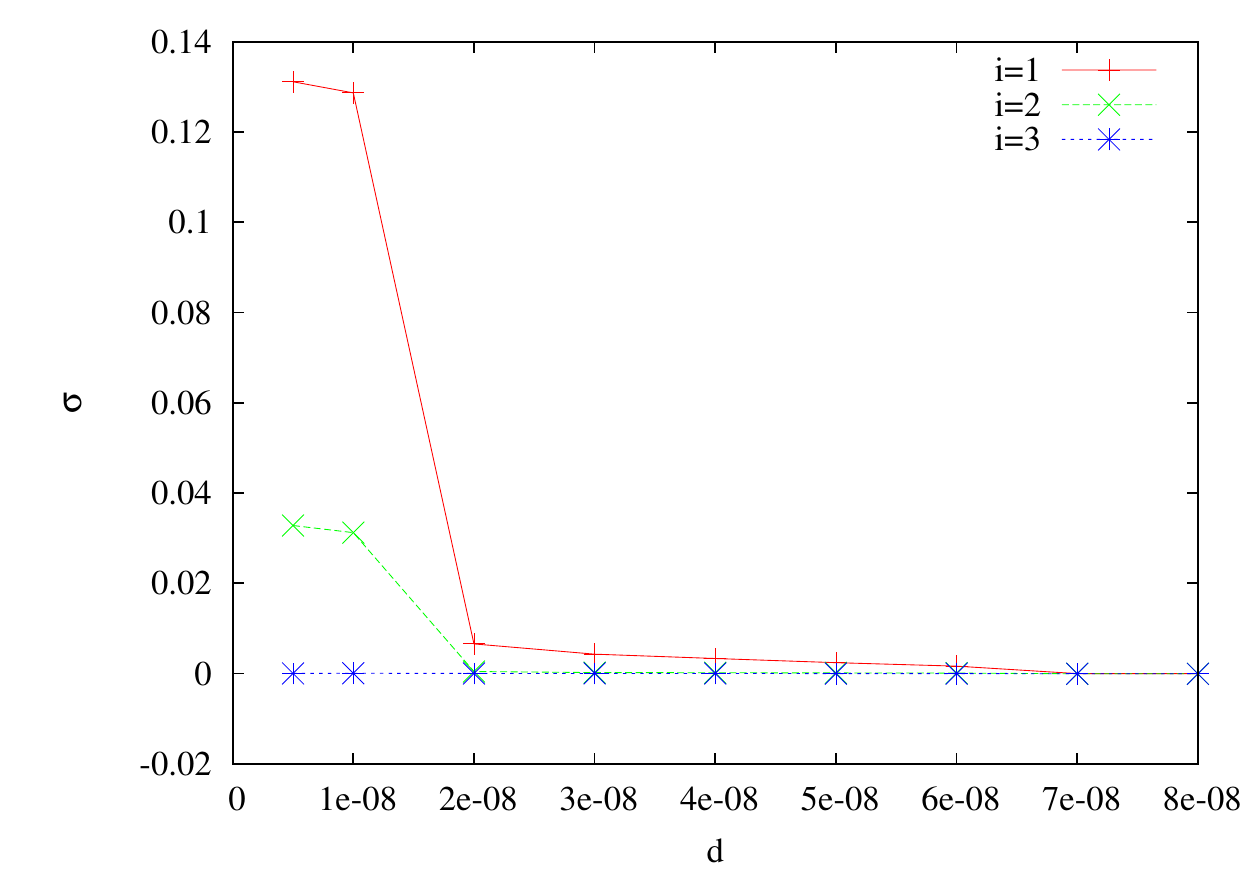}}
{\includegraphics[width=100mm]{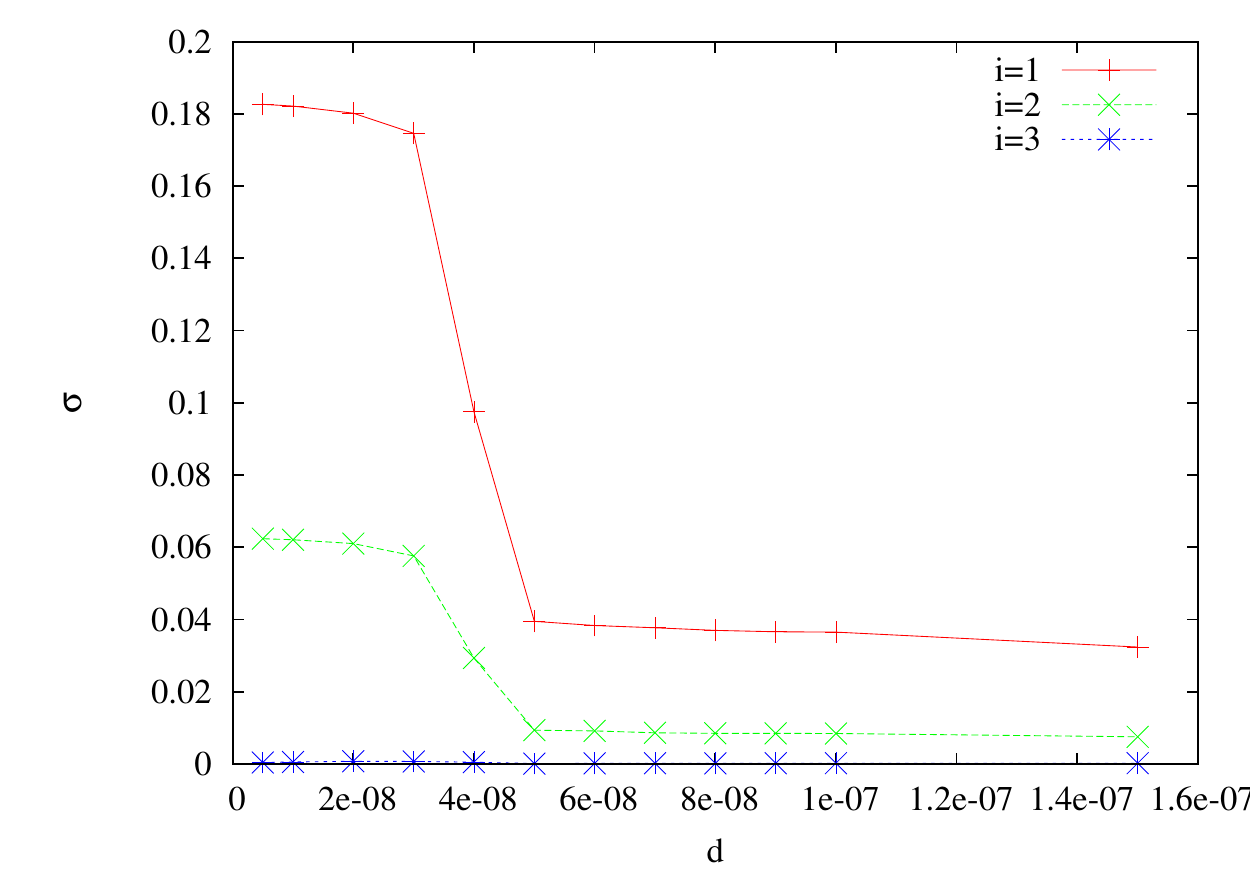}}
\caption{
Leading Lyapunov exponents as a function of $d$. In both panels $C_{\rm a}=C_{\rm o}/4$ and $\lambda = 20$~Wm$^{-2}$s$^{-1}$:
(a) $C_{\rm o} = 300$ Wm$^{-2}$, and (b) $C_{\rm o}=350$ Wm$^{-2}$; $d$-values on the abscissa have been multiplied by $10^8$ [s$^{-1}$]. Otherwise same as Fig.~\ref{fig:lead_Lyap}. 
} 
\label{fig:lead_Lyap_d}
\end{figure}

\begin{figure}
\centering
{\includegraphics[width=100mm]{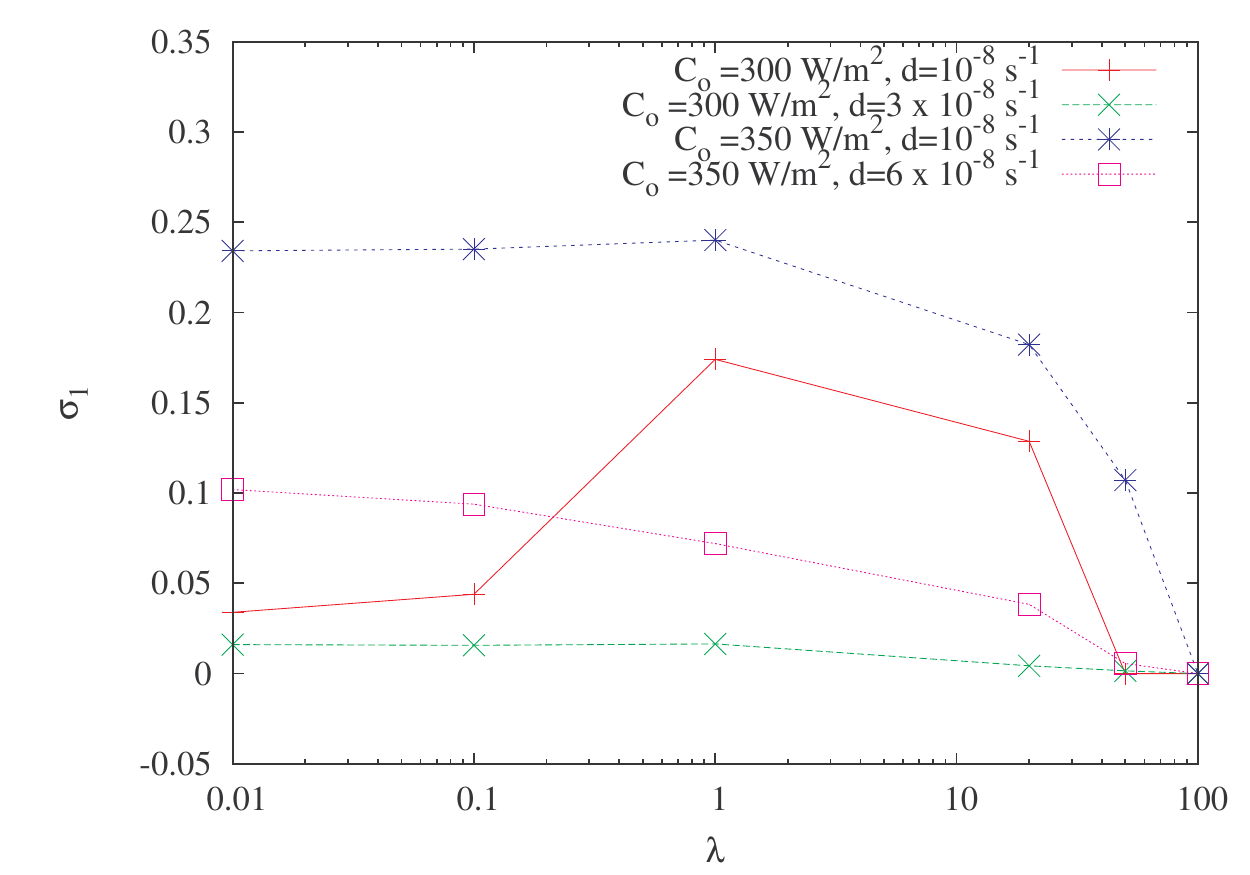}}
\caption{First Lyapunov exponent $\sigma_1$ as a function of the heat flux parameter $\lambda$
	[Wm$^{-2}$K$^{-1}$], for different combinations of $(C_{\rm o}, d)$. 
	}
\label{fig:LyaP_heat}
\end{figure}

\begin{figure}
\centering
{\includegraphics[width=100mm]{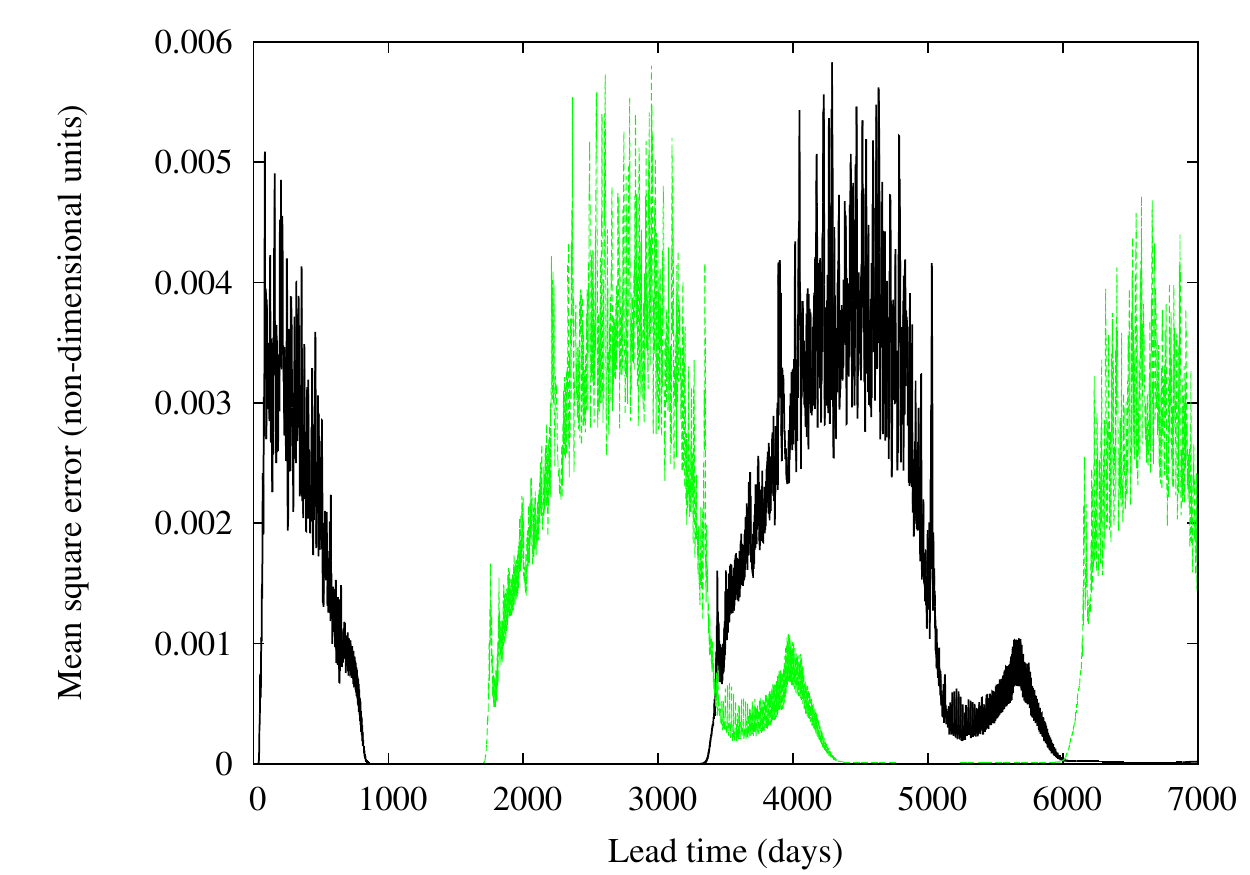}}
\caption{Root-mean-square evolution of the errors in the model's atmospheric variables, 
as averaged over an ensemble of 1000 realizations, starting from perturbations of two distinct initial states on the coupled model's 
     attractor, for $C_{\rm o}=350$ Wm$^{-2}$ and $d=6 \times 10^{-8} s^{-1}$; 
     distribution of mean 0 and variance $10^{-12}$ (in adimensional units) affecting all the variables of the coupled model.
     see text for details.The two initial states we have perturbed for the two ensembles are indicated 
     in Fig.~\ref{fig:KS} by the filled black circle for the black curve and the green triangle for the green 
     curve, respectively.
     }
\label{fig:predict}
\end{figure}

\end{document}